\documentclass[twocolumn,trackchanges]{aastex701}

\usepackage{amsmath}
\shorttitle{CS survey of disks}
\shortauthors{Law et al.}

\begin{document}

\title{A Submillimeter Survey of CS Excitation in Protoplanetary Disks: Evidence of X-ray-Driven Sulfur Chemistry}

\author[0000-0003-1413-1776]{Charles J.\ Law}
\altaffiliation{NASA Hubble Fellowship Program Sagan Fellow}
\affiliation{Department of Astronomy, University of Virginia, Charlottesville, VA 22904, USA}
\email[show]{cjl8rd@virginia.edu}  

\author[0000-0003-1837-3772]{Romane Le Gal}
\affiliation{Université Grenoble Alpes, CNRS, IPAG, F-38000 Grenoble, France}
\affiliation{Institut de Radioastronomie Millimetrique (IRAM), 300 rue de la piscine, F-38406 Saint-Martin d’Hères, France}
\email{romane.le-gal@univ-grenoble-alpes.fr}

\author[0000-0001-8798-1347]{Karin I. \"Oberg}
\affiliation{Center for Astrophysics \textbar\, Harvard \& Smithsonian, 60 Garden St., Cambridge, MA 02138, USA}
\email{koberg@cfa.harvard.edu}

\author[0000-0002-0661-7517]{Ke Zhang}
\affiliation{Department of Astronomy, University of Wisconsin-Madison, 475 N Charter St, Madison, WI 53706}
\email{ke.zhang@wisc.edu}

\author[0000-0003-3283-6884]{Yuri Aikawa}
\affiliation{Department of Astronomy, The University of Tokyo, 7-3-1 Hongo, Bunkyo-ku, Tokyo 113-0033, Japan}
\email{aikawa@astron.s.u-tokyo.ac.jp}

\author[0000-0003-2253-2270]{Sean M. Andrews}
\affiliation{Center for Astrophysics \textbar\, Harvard \& Smithsonian, 60 Garden St., Cambridge, MA 02138, USA}
\email{sandrews@cfa.harvard.edu}

\author[0000-0001-7258-770X]{Jaehan Bae}
\affiliation{Department of Astronomy, University of Florida, Gainesville, FL 32608, USA}
\email{jbae@ufl.edu}

\author[0000-0003-2014-2121]{Alice S. Booth} 
\altaffiliation{Clay Postdoctoral Fellow}
\affiliation{Center for Astrophysics \textbar\, Harvard \& Smithsonian, 60 Garden St., Cambridge, MA 02138, USA}
\email{alice.booth@cfa.harvard.edu}

\author[0000-0002-2700-9676]{Gianni Cataldi}
\affiliation{National Astronomical Observatory of Japan, Osawa 2-21-1, Mitaka, Tokyo 181-8588, Japan}
\email{gianni.cataldi@nao.ac.jp}

\author[0000-0003-2076-8001]{L. Ilsedore Cleeves}
\affiliation{Department of Astronomy, University of Virginia, Charlottesville, VA 22904, USA}
\email{lic3f@virginia.edu}

\author[0000-0002-7607-719X]{Feng Long}
\altaffiliation{NASA Hubble Fellowship Program Sagan Fellow}
\affiliation{Lunar and Planetary Laboratory, University of Arizona, Tucson, AZ 85721, USA}
\email{fenglong@arizona.edu}

\author[0000-0002-1637-7393]{François Ménard}
\affiliation{Université Grenoble Alpes, CNRS, IPAG, F-38000 Grenoble, France}
\email{francois.menard@univ-grenoble-alpes.fr}

\author[0000-0001-8642-1786]{Chunhua Qi}
\affiliation{Institute for Astrophysical Research, Boston University, 725 Commonwealth Avenue, Boston, MA 02215, USA}
\email{qi.molecules@gmail.com}

\author[0000-0003-1534-5186]{Richard Teague}
\affiliation{Department of Earth, Atmospheric, and Planetary Sciences, Massachusetts Institute of Technology, Cambridge, MA 02139, USA}
\email{rteague@mit.edu}

\author[0000-0003-1526-7587]{David J. Wilner}
\affiliation{Center for Astrophysics \textbar\, Harvard \& Smithsonian, 60 Garden St., Cambridge, MA 02138, USA}
\email{dwilner@cfa.harvard.edu}

\begin{abstract}
The sulfur chemistry in protoplanetary disks influences the properties of nascent planets, including potential habitability. Although the inventory of sulfur molecules in disks has gradually increased over the last decade, CS is still the most commonly-observed sulfur-bearing species and it is expected to be the dominant gas-phase sulfur carrier beyond the water snowline. Despite this, few dedicated multi-line observations exist, and thus the typical disk CS chemistry is not well constrained. Moreover, it is unclear how that chemistry~--~and in turn, the bulk volatile sulfur reservoir~--~varies with stellar and disk properties. Here, we present the largest survey of CS to date, combining both new and archival observations from ALMA, SMA, and NOEMA of 12~planet-forming disks, covering a range of stellar spectral types and dust morphologies. Using these data, we derived disk-integrated CS gas excitation conditions in each source. Overall, CS chemistry appears similar across our sample with rotational temperatures of ${\approx}$10-40~K and column densities between 10$^{12}$-10$^{13}$~cm$^{-2}$. CS column densities do not show strong trends with most source properties, which broadly suggests that CS chemistry is not highly sensitive to disk structure or stellar characteristics. We do, however, identify a positive correlation between stellar X-ray luminosity and CS column density, which indicates that the dominant CS formation pathway is likely via ion-neutral reactions in the upper disk layers, where X-ray-enhanced S$^+$ and C$^+$ drive abundant CS production. Thus, using CS as a tracer of gas-phase sulfur abundance requires a nuanced approach that accounts for its emitting region and dependence on X-ray luminosity.
\end{abstract}



\section{Introduction} \label{sec:intro}

The properties of planetary systems are closely linked to the gas composition and physical conditions of their natal protoplanetary disks. Sensitive sub-millimeter observations, such as those from the Atacama Large Millimeter/submillimeter Array (ALMA), have enabled detailed characterization of the molecular gas inventory, distribution, and excitation in increasing numbers of disks \citep[e.g.,][]{Thi04, Dutrey11, Guilloteau13, Rosenfeld13, Bergner19, Miotello19, Pegues20, Loomis20, Oberg21_MAPSI, Law21_MAPSIII, Phuong21, Stapper24, Booth24_HD100546}. Much of this work has, to date, targeted species containing carbon, oxygen, and nitrogen, while comparatively fewer studies have surveyed sulfur-bearing molecules \citep[e.g.,][]{Fuente10, Guilloteau16, Phuong18, Semenov18, LeGal19, Legal21, Booth24_IRS}. As a result, the gas-phase sulfur chemistry and the underlying sulfur reservoir remain poorly understood in disks.

The uncertainty in the sulfur budget of planet-forming disks reflects long-standing open questions about the chemical partitioning of sulfur across astrophysical environments. For instance, sulfur abundances are approximately solar \citep[S/H~$\sim 1.5\times$10$^{-5}$;][]{Asplund09} in the diffuse interstellar medium \citep{Howk06} and photodissociation regions \citep{Goicoechea06}, but are strongly depleted in dense molecular gas \citep[e.g.,][]{Tieftrunk94, Vastel18}, including in protoplanetary disks. Disks show volatile sulfur depletion factors up to ${\sim}$1000 \citep{Semenov18, Legal21, Keyte24_HD169142, Keyte24_HD100546} with much of this missing sulfur potentially locked up in ices \citep[e.g.,][]{Millar90,Ruffle99,Vidal17,Laas19,Booth21_IRS} or in refractory sulfide minerals such as FeS \citep{Kama19}. 

Complementary insights come from cometary studies, where H$_2$S is the major volatile sulfur-bearing species \citep{Calmonte16}. However, detections of gas-phase H$_2$S in disks remain scare \citep[e.g.,][]{Dutrey11, Phuong18, Semenov18, RM_ABaur_H2S, Keyte24_HD100546}, while H$_2$S remains undetected in interstellar ices in dense clouds \citep{Smith91, McClure23} and young stellar objects \citep{JimEso_Munoz11, Boogert15}. Moreover, elevated H$_2$CS/CS gas-phase column density ratios measured in several disks around Herbig stars \citep{LeGal19, Legal21, Booth23_HD169142, Booth24_HD100546, Law25} are in tension with chemical models and suggest that organic sulfur compounds may comprise a large fraction of the volatile sulfur reservoir. While several complex and mutli-sulfuretted molecules (e.g., S$_2$, S$_3$, S$_4$, CH$_3$SH, C$_2$H$_6$S) were also detected in Comet 67P, together they only contributed a few percent of the total sulfur budget \citep{Calmonte16}.  Overall, this indicates that sulfur chemistry -- and the volatile and refractory sulfur reservoirs -- is far from understood in disks, with significant implications for the chemical origins of our own Solar System and exoplanetary systems. This is especially critical as sulfur is thought to play a key role in driving prebiotic chemistry and determining planet habitability \citep[][]{Leman04, Patel15, Ranjan18}, while the recent detection of sulfur-bearing molecules in exoplanetary atmospheres \citep[e.g.,][]{Powell24, Dyrek24, Beatty24} may provide a powerful link between disk and planetary compositions if disk sulfur chemistry is more well understood. 

Over the last decade, the inventory of gas-phase sulfur molecules in disks has gradually increased with  detections of new organic \citep[e.g., H$_2$CS, H$_2$S, CCS;][]{Phuong18, LeGal19, Loomis20, Phuong21}, oxygenated \citep[SO$_2$;][]{Booth21_IRS}, and $^{33}$S and $^{34}$S isotopologues \citep[e.g., C$^{34}$S, $^{33}$SO, $^{34}$SO, $^{34}$SO$_2$, H$_2$C$^{34}$S;][]{LeGal19, Loomis20, Booth24_HD100546, Booth24_IRS, Booth24_S_iso, Law25}. However, many of these species have only been detected in a few disks and require deep observations due to their faint line emission. As a result, the most commonly detected sulfur-bearing species remains the CS molecule, which has now been observed in dozens of disks \citep[e.g.,][]{Dutrey97,vanderPlas14,Guilloteau16, Semenov18, LeGal19, Legal21, Rosotti21, Facchini21PDS, Huang24, Law25, Teague25}. In most disks, CS is expected to be the dominant gas-phase sulfur carrier \citep[e.g.,][]{Legal21} at large disk radii, i.e., beyond the water sublimation front, and thus is an ideal tracer to probe the bulk sulfur content. Due to its relatively bright lines, CS is also readily detectable in shorter integrations and thus, widely accessible across larger disk samples. Despite this, few dedicated multi-line observations exist, which means that CS excitation conditions have only been robustly determined in a small number of disks \citep[e.g.,][]{Teague18_TWHya, LeGal19, Law25}. Thus, the typical disk CS chemistry is not well constrained and it is unclear how this chemistry -- and in turn, the bulk volatile sulfur chemistry -- varies depending on disk and stellar properties. 

In this paper, we present a comprehensive set of multi-line CS data in a sample of 12 protoplanetary disks compiled from new and archival observations. Using these data, we determine rotational temperatures and column densities for each disk in our sample, which we use to assess the nature of CS chemistry in disk environments. In Section \ref{sec:disk_sample}, we briefly describe our disk sample and summarize the observations used in this work in Section \ref{sec:observations_overview}. We present our results in Section \ref{sec:results}. In Section \ref{sec:discussion}, we explore how sulfur chemistry varies across our sample and discuss the chemical origins of CS in planet-forming disks. We summarize our conclusions in Section \ref{sec:conlcusions}.

\setlength{\tabcolsep}{2.5pt}
\begin{deluxetable*}{lcccccccccc}
\tablecaption{Stellar and Disk Properties of our Sample\label{tab:source_prop}}
\tablewidth{0pt}
\tablehead{
\colhead{Source} & \colhead{Region} & \colhead{SpT}  &  \colhead{Dist.$^{[2]}$} & \colhead{M$_*$\tablenotemark{a}} & \colhead{L$_*$} & \colhead{Age} & \colhead{v$_{\rm{sys}}$\tablenotemark{a}} & \colhead{incl.} & \colhead{P.A.} & \colhead{L$_{\rm{X}}$\tablenotemark{d}} \\ 
\colhead{}  & \colhead{} & \colhead{} & \colhead{(pc)}   & \colhead{(M$_{\odot}$)} & \colhead{(L$_{\odot}$)} & \colhead{(Myr)} & \colhead{(km~s$^{-1}$)} &  \colhead{($^{\circ}$)} & \colhead{($^{\circ}$)} & \colhead{(10$^{30}$ erg s$^{-1}$)} 
}
\startdata
IM~Lup & Lupus & K5$^{[1]}$ & 158 & 1.1$^{[3]}$ & 2.6$^{[2]}$ & ${\sim}$1$^{[4]}$ & 4.5$^{[3]}$ & 47.5$^{[5]}$ & 144.5$^{[5]}$ & 0.96$^{[39]}$ \\ 
GM~Aur & Taurus & K6$^{[6]}$ & 159 & 1.1$^{[3]}$ & 1.2$^{[7]}$ & ${\sim}$3-10$^{[8]}$ & 5.6$^{[3]}$ & 53.2$^{[9]}$ & 57.2$^{[9]}$ & 4.7$^{[40]}$ \\ 
AS~209 & Oph~N~3a & K5$^{[10]}$ & 121 & 1.2$^{[3]}$ & 1.4$^{[10]}$ & ${\sim}$1-2$^{[10]}$ & 4.6$^{[3]}$ & 35.0$^{[5]}$ & 85.8$^{[5]}$ & 4.1$^{[41]}$ \\
HD~163296 & isolated? & A1$^{[11]}$ & 101 & 2.0$^{[3]}$ & 17.0$^{[11]}$ & ${\gtrsim}6$$^{[11]}$ & 5.8$^{[3]}$ & 46.7$^{[5]}$ & 133.3$^{[5]}$ & 0.47$^{[42]}$ \\
MWC~480 & Taurus & A5$^{[12]}$ & 162 & 2.1$^{[3]}$ & 21.9$^{[12]}$ & ${\sim}$7$^{[12]}$ & 5.1$^{[3]}$ & 37.0$^{[13]}$ & 148.0$^{[13]}$ & 0.81$^{[43]}$\\ 
DM~Tau & Taurus & M1$^{[14]}$ & 145 & 0.5$^{[15]}$ & 0.24$^{[16]}$ & ${\sim}$3-7$^{[16]}$ & 6.0$^{[15]}$ & 35.2$^{[17]}$ & 157.8$^{[17]}$ & 2.5$^{[44]}$ \\
GG~Tau\tablenotemark{c} & Taurus & M0+M2,~M3$^{[18]}$ & 150 & 1.2$^{[18]}$ & 0.67$^{[19]}$ & ${\sim}$3$^{[18]}$ & 6.4$^{[20]}$ & 35.0$^{[21]}$ & 7.0$^{[21]}$ & 0.29$^{[45]}$ \\
LkCa~15 & Taurus & K5$^{[22]}$ & 157 & 1.2$^{[15]}$ & 1.1$^{[22]}$ & ${\sim}$5$^{[22]}$ & 6.3$^{[15]}$ & 50.2$^{[23]}$ & 61.9$^{[23]}$ & 2.0$^{[46]}$ \\
V4046~Sgr\tablenotemark{c} & $\beta$ Pic MG & K5+K7$^{[24]}$ & 73 & 1.8$^{[25]}$ & 0.86$^{[26]}$ & ${\sim}$12-23$^{[27]}$ & 2.9$^{[26]}$ & 34.7$^{[26]}$  & 75.7$^{[26]}$ & 1$^{[47]}$  \\
J1604\tablenotemark{b} & Upper Sco & K2$^{[28]}$ & 145 & 1.2$^{[29]}$ & 0.6$^{[30]}$ & ${\sim}$5-10$^{[31]}$ & 4.6$^{[29]}$ & 6.0$^{[31]}$ & 258.8$^{[29]}$ & \ldots  \\
CI~Tau & Taurus & K5.5$^{[32]}$ & 160 & 1.0$^{[33]}$ & 1.26$^{[34]}$ & ${\sim}$2$^{[34]}$ & 5.7$^{[33]}$ & 49.2$^{[35]}$ & 11.3$^{[35]}$ & 0.15$^{[48]}$\\ 
TW~Hya & TWA & K7$^{[36]}$ & 60 & 0.81$^{[37]}$ & 0.23$^{[38]}$ & ${\sim}$8-10$^{[38]}$ & 2.8$^{[37]}$ &  5.8$^{[37]}$ & 151.6$^{[37]}$ & 1.4$^{[49]}$
\enddata
\tablecomments{References are: 1.~\citet{Alcala17}; 2.~\citet{Gaia23}; 3.~\citet{Teague21}; 4.~\citet{Mawet12}; 5.~\citet{Huang18}; 6.~\citet{Espaillat10}; 7.~\citet{Macias18}; 8.~\citet{Kraus09}; 9.~\citet{Huang20}; 10.~\citet{Andrews18}; 11.~\citet{Fairlamb15}; 12.~\citet{Montesinos09}; 13.~\citet{Liu19}; 14.~\citet{Kenyon95}; 15.~\citet{Law23}; 16.~\citet{Pegues20}; 17.~\citet{Kudo18}; 18.~\citet{Keppler20}; 19.~\citet{Hartigan03}; 20.~\citet{Guilloteau99}; 21.~\citet{Phuong21}; 22.~\citet{Donati19}; 23.~\citet{Facchini20}; 24.~\citet{Quast00}; 25.~\citet{Flaherty20}; 26.~\citet{Rosenfeld12}; 27.~\citet{Mamajek14}; 28.~\citet{Preibisch05}; 29.~\citet{Stadler23}; 30.~\citet{carpenter_alma_2014}; 31.~\citet{dong_sizes_2017}; 32.~\citet{Simon19}; 33.~\citet{Law22}; 34.~\citet{Donati20}; 35.~\citet{Clarke18}; 36.~\citet{Ingleby13}; 37.~\citet{Teague19_TWHya}; 38.~\citet{Sokal18}; 39.~\citet{Cleeves17}; 40.~\citet{Espaillat19}; 41.~\citet{Walter81}; 42.~\citet{Gunther09}; 43.~\citet{Anilkumar24}; 44.~\citet{Long24}; 45.~\citet{Bary03}; 46.~\citet{Skinner17}; 47.~\citet{Gunther06}; 48.~\citet{Gudel07}; 49.~\citet{Argiroffi17}. \\
\tablenotetext{a}{Dynamical stellar masses and systemic velocities (in the LSR frame) are derived via resolved observations of CO isotopologue lines.}
\tablenotetext{b}{We hereafter refer to this source, 2MASS J16042165-2130284 (also known as RXJ1604.3-2130~A), as J1604 throughout the text.}
\tablenotetext{c}{For GG~Tau and V4046~Sgr, the stellar mass and bolometric luminosity are the sum of all stars in each system.}
\tablenotetext{d}{We report the most recent stellar X-ray luminosity available, but note that several systems are X-ray active and experience transient increases in L$_{\rm{X}}$ by factors of several during flaring events \citep[e.g.,][]{Cleeves17, Espaillat19}. All L$_{\rm{X}}$ values are scaled to the reported source distances.}}
\end{deluxetable*}
\setlength{\tabcolsep}{4pt}

\section{Disk Sample} \label{sec:disk_sample}

Table \ref{tab:source_prop} summarizes the properties of the 12 sources in our survey, which consist of well-studied disks around both T~Tauri and Herbig~Ae stars. Each star hosts a radially-extended (100s of au), gas-rich disk \citep[e.g.,][]{Huang18_TWHya, Phuong20, Law21_MAPSIII, Rosotti21, Law23, Stadler23} with diverse dust substructures, including rings, gaps, spiral arms, and non-axisymmetric features \citep[e.g.,][]{Huang18, Kudo18, Facchini20, Sierra21, Long22_LkCa15, Curone25}. We note that this is not an unbiased survey but instead comprises many large and bright disks that have been the subject of extensive chemical observations, including the five disks surveyed as part of the Molecules with ALMA at Planet-forming Scales (MAPS) ALMA Large Program \citep{Oberg21_MAPSI}. Our sample includes several transition disks (DM~Tau, GM~Aur, J1604, LkCa~15, TW~Hya) and the circumbinary disks around GG Tau and V4046~Sgr, which both have central dust-depleted cavities. The radial extent of these inner cavities range from a few au (TW~Hya) to over 200~au (GG~Tau). Stellar hosts range from M- to A-type with masses and bolometric luminosities of 0.5-2.1~M$_{\odot}$ and 0.23-21.9~L$_{\odot}$, respectively. All systems have existing dynamical stellar masses derived via resolved observations of Keplerian gas rotation \citep[e.g.,][]{Teague21}. Across our sample, stellar X-ray luminosities span almost two orders of magnitude from ${\approx}$10$^{29}$ to ${\approx}$5$\times$10$^{30}$~erg~s$^{-1}$. Typical system ages are ${\sim}$1-10~Myr with the exception of V4046~Sgr, which may be a factor of a few older. All disks are nearby with distances ranging from 60~pc to 162~pc.

\section{Observations}
\label{sec:observations_overview}

We compiled published, archival, and new observations of planet-forming disks with at least two transitions of CS from the Submillimeter Array (SMA)\footnote{The Submillimeter Array is a joint project between the Smithsonian Astrophysical Observatory and the Academia Sinica Institute of Astronomy and Astrophysics and is funded by the Smithsonian Institution and the Academia Sinica.}, Northern Extended Millimeter Array (NOEMA), and ALMA. Below, we describe the observational details, data reduction, and typical angular and velocity resolution of the observations.

\subsection{SMA Observations and Imaging}

\textit{Projects 2020A-S018 and 2020B-S025}: We observed the IM~Lup, GM~Aur, AS~209, HD~163296, and MWC~480 disks with the SMA between 10 Aug 2020 and 28 May 2021 as part of projects 2020A-S018 and 2020B-S025 (PI: R. Le Gal). All sources were observed in the compact (COM) configuration of the SMA, with the exception of GM~Aur, which used the sub-compact (SUB-COM) configuration. The SUB-COM and COM observations had typical maximum baselines of ${\approx}$30~m and ${\approx}$70~m, respectively. All observations used either 7 or 8 antennas with zenith optical depths at 225 GHz ($\tau_{225\rm{GHz}}$) ranging from 0.05-0.16 as measured by the SMA's radiometer\footnote{\url{https://lweb.cfa.harvard.edu/sma/memos/164.pdf}}. Table \ref{tab:calb_details} shows a summary of the observations, including observing dates, SMA configuration, and number of available antennas.

\setlength{\tabcolsep}{3.0pt}
\begin{deluxetable*}{lccccccccc}
\tablecaption{SMA Observational Details\label{tab:calb_details}}
\tablewidth{0pt}
\tablehead{
\colhead{Source} & \colhead{R.A.} & \colhead{Dec.} & \colhead{UT Date} & \colhead{Num.} & \colhead{SMA} & \colhead{$\tau_{\rm{225GHz}}$\tablenotemark{b}} & \multicolumn3c{Calibrators} \\ \cline{8-10}
\colhead{} & \colhead{(J2000)} & \colhead{(J2000)} & \colhead{} & \colhead{Ant.\tablenotemark{a}} & \colhead{config.} &\colhead{} & \colhead{Flux} & \colhead{Passband} & \colhead{Gain}
}
\startdata
\multicolumn9c{2023A-S019 (PI: C. Law)} \\ \hline \hline
GM~Aur &  04:55:11.0 & $+$30:21:59.0 & 2023 Nov 05 & 6 & COM & 0.04 & Titan & 3C84 & 0510+180, 3C111 \\
       &  & & 2023 Oct 22 & 7 & EXT & 0.07  & Callisto & 3C84  & 0510+180, 3C111 \\
J1604  & 16:04:21.6 & $-$21:30:28.9 & 2023 Jun 26 & 7 & COM & 0.05 & Ceres & 3C454.3 & 1507-168, 1517-243 \\
LkCa~15 & 04:39:17.8 & $+$22:21:03.1 & 2023 Nov 06 & 6 & COM & 0.03 & Titan  & 3C84  & 0510+180, 3C111 \\
       &  & & 2023 Oct 20 & 7 & EXT & 0.12 & Titan & 3C84 & 0510+180, 3C111  \\
GG~Tau & 04:32:30.3 & $+$17:31:40.6 & 2023 Nov 12 & 7 & COM & 0.11 & Uranus & 3C84  & 0510+180, 3C120 \\
       &  & & 2023 Oct 30 & 7  & EXT & 0.09  & Titan  &  3C84 & 0510+180, 3C120 \\
V4046~Sgr  & 18:14:10.5 & $-$32:47:35.3 & 2023 Jul 14 & 7 & COM & 0.06 & Callisto & 3C279 & 1700-261, NRAO~530  \\ \hline 
\multicolumn9c{2020A-S018 + 2020B-S025  (PI: R. Le Gal)} \\ \hline
\hline 
IM~Lup & 15:56:09.2  & $-$37:56:06.1 & 2021 May 28 & 7 & COM & 0.12 & Callisto & 3C279 & 1517-243, 1626-298\\
GM~Aur & 04:55:10.0 & $+$30:21:59.0 & 2020 Nov 02 & 8 & SUB-COM & 0.05 & Uranus & 3c454.3 & 3C111, 0510+180 \\
AS~209 & 16:49:15.0 & $-$14:22:08.0 & 2020 Aug 17 & 8 & COM & 0.15 & Uranus & 3C454.3 & 1733-130 \\
HD~163296 & 17:56:21.0 & $-$21:57:21.0 & 2020 Aug 10 & 8\tablenotemark{c} & COM & 0.16 & Uranus & 3C454.3 & 1743-038, NRAO~530 \\
MWC~480 & 04:58:46.0 & $+$29:50:36.0 & 2020 Aug 10 & 8 & COM & 0.16 & Uranus & 3C454.3 & 3C111, 0510+180\\
\enddata
\tablenotetext{a}{Number of antennas remaining after flagging.}
\tablenotetext{b}{Zenith optical depth at 225~GHz measured with the SMA's radiometer.}
\tablenotetext{c}{Due to unstable phases, data from Antenna 8 were flagged, resulting in 7 total antennas in the 240~GHz receiver data that covers the CS J=5--4 line.}
\end{deluxetable*}
\setlength{\tabcolsep}{4pt}

All observations used both the 230~GHz and 240~GHz receivers and the SWARM correlator, which at the time, provided 32~GHz of bandwidth. The two sidebands of the lower frequency receiver covered 194.9-202.9 GHz and 210.9-218.9 GHz, while the higher frequency receiver was tuned to span 227.0-231.0~GHz and 241.0-249.0~GHz. The tuning selection was motivated by the desire to simultaneously cover the CS J=4--3 (195.95421~GHz) and CS J=5--4 (244.93564~GHz) lines. The native spectral resolution of the observations was 140~kHz (${\approx}$0.2~km~s$^{-1}$), but was later binned to a uniform velocity resolution of 1~km~s$^{-1}$ during imaging.

We calibrated the SMA data using the MIR software package\footnote{\url{https://lweb.cfa.harvard.edu/~cqi/mircook.html}}. Passband calibrations were performed using observations of the bright quasars 3C279 or 3C454.3. Depending on the source, flux calibration was achieved using either Callisto or Uranus, while several quasars were used as gain calibrators. Table \ref{tab:calb_details} shows a detailed listing of all calibrators. The calibrated visibilities were then exported into CASA \texttt{v6.3} \citep{McMullin_etal_2007, CASA22} for continuum subtraction and imaging. 

\textit{Project 2023A-S019}: We also observed the GM~Aur, GG~Tau, LkCa~15, V4046~Sgr, and J1604 disks between 26 June 2023 and 12 Nov 2023 as part of the SMA project 2023A-S019 (PI: C. Law). All sources were observed in the COM configuration, while GM~Aur, GG~Tau, and LkCa~15 also had data taken in the extended (EXT) configuration, which had baselines up to ${\approx}$220~m. Each observation used either 6 or 7 antennas with $\tau_{225\rm{GHz}}$ between 0.03-0.12. Table \ref{tab:calb_details} provides a summary of the observations.

Each observation used the 240~GHz and 345~GHz receivers and the upgraded SWARM correlator, which provided 48~GHz of bandwidth. The lower frequency receiver covered 244.3-256.6~GHz and 264.3-276.6~GHz, while the higher frequency receiver included 333.8-346.1~GHz and 353.8-366.1~GHz. While this tuning was originally selected to probe lines of HNC, which will be presented in a forthcoming publication, the wide spectral coverage of the SMA also covered the CS J=5--4 (244.93564~GHz) and CS J=7--6 (342.88285~GHz) lines.

We converted the raw data to CASA measurement sets using the \texttt{pyuvdata} \citep{Hazelton17} SMA reduction pipeline\footnote{\url{ https://github.com/Smithsonian/sma-data-reduction}} in CASA version \texttt{v6.3}. Depending on the source and configuration, we used 3C84, 3C454.3, or 3C279 for our passband calibrators; Titan, Callisto, Ceres, or Uranus as flux calibrators; and several quasars as gain calibrators. Table \ref{tab:calb_details} lists all calibrators used. To reduce data volume, we binned by a factor of four for initial calibration, followed by a further smoothing to a uniform velocity resolution of 1~km~s$^{-1}$ during imaging.

For data from both SMA programs, we first subtracted the continuum using the \texttt{uvcontsub} task in CASA with a first-order polynomial. We then imaged all lines using the \texttt{tclean} task with Briggs weighting and Keplerian masks generated with the \texttt{keplerian\_mask} \citep{rich_teague_2020_4321137} code. We chose consistent Keplerian mask parameters for each disk that matched the known stellar properties and disk geometry from the literature. We used natural weighting (\texttt{robust}=2) and binned to 1~km~s$^{-1}$ channels to prioritize high SNR detections while still well-sampling the disk-integrated line full-width at half maximum (FWHM). All images were CLEANed down to a 3$\sigma$ level, where $\sigma$ was the RMS noise measured across ten line-free channels of the dirty image. For GM~Aur, GG~Tau, and LkCa~15, which were observed in two configurations, we combined all data prior to imaging. We also occasionally applied an additional \textit{uv}-taper to a few lines to increase the SNR of the resulting images. Across the SMA disk sample, typical synthesized beam sizes were ${\approx}$2-5$^{\prime \prime}$ and line RMS noise values ranged from several tens of mJy~beam$^{-1}$ in the lower frequency CS J=4--3 and 5--4 lines to ${\gtrsim}$100~mJy~beam$^{-1}$ in the higher frequency CS J=7--6 line. Table \ref{tab:obs-list} summarizes all image cube properties.

\setlength{\tabcolsep}{2.5pt}
\begin{table*}[!]
\scriptsize
\begin{center}
\caption{CS Image Cube Properties \label{tab:obs-list}}
 \renewcommand{\arraystretch}{1.2}
\begin{tabular}{lccccllccccc}
\hline\hline
Species&Line&Freq.&E$_{\rm{u}}$&A$_{\rm{ul}}$&Source& $\delta$V& RMS & Beam&R$_{\rm{max}}^{(a)}$&$S_\nu\Delta_v$(R$_{\rm{max}}^{(b)}$) & Ref. \\
&&(GHz)&(K)&($\log_{10}$ s$^{-1}$)&&(km~s$^{-1}$)&(mJy~beam$^{-1}$)&($''\, \times \,''$, \degr)&($^{\prime \prime}$)&(mJy~km~s$^{-1}$)  \\ 
\hline
CS& J=$2-1$&97.98095 & 7.1 & $-$4.776 & IM~Lup & 0.22 & 0.64 & $0.3\times0.3$,~0 & 4.0 & 341~$\pm$~16 & 1,2 [ALMA] \\ 
&&&&& GM~Aur & 0.22 & 1.26 & $0.3\times0.3$,~0 & 3.0 & 271~$\pm$~11 & 1,2 [ALMA] \\
&&&&& AS~209 & 0.22 & 0.65 & $0.3\times0.3$,~0 & 1.8 & 165~$\pm$~9 & 1,2 [ALMA] \\ 
&&&&& MWC~480 & 0.22 & 1.27 & $0.3\times0.3$,~0 & 2.0 & 56~$\pm$~11 & 1,2 [ALMA] \\
&&&&& GG~Tau & 0.19 & 4.0 & 4.2~$\times$~3.2,~1.0 & 7.0 & 286~$\pm$~5 & 3 [NOEMA] \\
\hline
& J=$3-2$ & 146.96903 & 14.1 & $-$4.218 & DM~Tau & 0.30 & 5.0 & 0.6~$\times$~0.5,~23 & 3.5 & 1100~$\pm$~300& 4 [ALMA]  \\
\hline
& J=$4-3$ & 195.95421 & 23.5 & $-$3.828 & IM~Lup & 1.0 & 65.1 & 5.7~$\times$~3.4,~$-$11.7 & 4.0 & 1071~$\pm$~116 & 5 [SMA] \\
&&&&& GM~Aur & 1.0 & 37.2 & 4.8~$\times$~3.9,~$-$17.1 & 3.0 & 1153~$\pm$~69 & 5 [SMA] \\
&&&&& AS~209 & 1.0 & 59.8 & 4.1~$\times$~3.6,~$-$11.3 & 1.8 & 427~$\pm$~54 & 5 [SMA] \\
&&&&& MWC~480 & 0.22 & 1.7 & 0.44~$\times$~0.44,~0.0 & 2.0 & 404~$\pm$~15 & 5 [ALMA] \\
\hline
& J=$5-4$ & 244.93564 &35.3&$-$3.527 & IM~Lup & 1.0 & 94.4 & 4.8~$\times$~2.8,~$-$13.8 & 4.0 & 1676~$\pm$~160 & 5 [SMA] \\
&&&&& GM~Aur & 1.0 & 30.2 & 4.8~$\times$~3.9,~$-$17.1 & 3.0 & 1334~$\pm$~51 & 5 [SMA] \\
&&&&& AS~209 & 1.0 & 83.2 & 3.3~$\times$~2.8,~27.8 & 1.8 & 985~$\pm$~96 & 5 [SMA]\\
&&&&& MWC~480 & 0.18 & 3.9 & $0.71\times0.45$,~$-$12.9 & 2.0 & 536~$\pm$~19 & 1,2 [ALMA]\\
&&&&& GG~Tau & 1.0 & 71.0 & 3.1~$\times$~3.0,~44.4 & 7.0 & 1274~$\pm$~279 & 5 [SMA] \\
&&&&& LkCa~15 & 0.5 & 2.8 & 0.47~$\times$~0.57,~20.8 & 3.8 & 1050~$\pm$~14 & 6 [ALMA]\\
&&&&& V4046~Sgr & 1.0 & 82.2 & 4.3~$\times$~2.5,~9.2 & 2.8 & 2144~$\pm$~372 & 5 [SMA] \\
&&&&& J1604 & 0.7 & 108.0 & 3.7~$\times$~2.6,~7.9 & 3.0 & 873~$\pm$~243 & 5 [SMA] \\
&&&&& DM~Tau & 0.5 & 3.1 & 0.47~$\times$~0.54,~20.4 &  3.5 & 388~$\pm$~9 & 6 [ALMA]  \\
&&&&& CI~Tau & 0.5 & 2.8 & 0.46~$\times$~0.57,~16.0 & 3.8 & 756~$\pm$~17 & 6 [ALMA] \\
\hline
& J=$6-5$ & 293.91224 & 49.4&$-$3.283 & MWC~480 & 1.0 & 8.8 & 0.36~$\times$~0.78,~$-$26.5 & 2.0 & 688~$\pm$~13 & 6 [ALMA] \\
&&&&& LkCa~15 & 1.0 & 8.3 & 0.38~$\times$~0.70,~$-$32.0 & 3.8 & 1282~$\pm$~25 & 6 [ALMA] \\
&&&&& V4046~Sgr & 0.1 & 6.2 & 0.67~$\times$~0.50,~$-$80.0 & 2.8 & 2608~$\pm$~52 & 6 [ALMA] \\
\hline
& J=$7-6$ & 342.88285 & 65.8 & $-$3.077 & GM~Aur & 1.0 & 80.6 & 2.9~$\times$~2.9,~$-$87.7 & 3.0 & 496~$\pm$~197 & 5 [SMA] \\
&&&&& GG~Tau & 1.0 & 128.3 & 2.9~$\times$~2.9,~74.4 & 7.0 & 487~$\pm$~441 & 5 [SMA] \\
&&&&& LkCa~15 & 1.0 & 79.4 & 3.0~$\times$~2.9,~48.1 & 3.8 & 641~$\pm$~309 & 5 [SMA] \\
&&&&& J1604 & 0.7 & 179.5 & 2.6~$\times$~1.9,~9.6 & 3.0 & 1181~$\pm$~414 & 5 [SMA] \\
&&&&& V4046~Sgr & 1.0 & 164.9 & 3.0~$\times$~1.9,~6.6 & 2.8 & 4103~$\pm$~841 & 5 [SMA] \\
&&&&& CI~Tau & 0.5 & 3.0 & 0.13~$\times$~0.10,~$-$35.1 & 3.8 & 746~$\pm$~36 & 7 [ALMA] \\
\hline
\end{tabular}
\end{center}
\tablenotetext{}{References: 1.~\citet{Legal21}; 2.~\citet{Oberg21_MAPSI}; 3.~\citet{Phuong21}; 4.~\citet{Semenov18}; 5.~this work; 6.~\citet{LeGal19}; 7.~\citet{Rosotti21}. Spectroscopic information for all lines are taken from the CDMS database \citep{Muller01, Muller05, Endres16}.}
\tablenotetext{a}{R$_{\rm{max}}$ indicates the outer radius of molecular line emission.}
\tablenotetext{b}{$S_\nu\Delta_v$(R$_{\rm{max}}$) corresponds to the flux density integrated out to R$_{\rm{max}}$.}
\end{table*}
\setlength{\tabcolsep}{4pt}

\subsection{ALMA Observations and Imaging}
\label{sec:ALMA_Obs}

Due to the faint CS emission in the MWC~480 disk, we do not detect the J=4--3 or J=5--4 lines in our SMA observations. For the latter, published ALMA observations with a CS J=5--4 detection exist (see next subsection). For the former lines, however, we make use of new ALMA Band 5 observations of the MWC~480 disk at ${\approx}$0\farcs4 from program 2021.1.00899.S (PI: K. Zhang). The data comprised four execution blocks (EBs) observed from 05 July 2022 to 07 August 2022 and used between 30-42 antennas for a total on-source integration time of 107~min. Projected baselines ranged from 15.1-2617.4~m and the maximum recoverable scale (MRS) was 3\farcs7 with a typical PWV of 0.7-1.7~mm. One spectral window was centered on the CS J=4--3 line with a resolution of 141.1~kHz (${\approx}$0.2~km~s$^{-1}$).

Following standard self-calibration procedures \citep[e.g.,][]{Andrews18}, we first re-aligned all individual execution blocks (EBs) to a common disk center using the \texttt{fixvis} and \texttt{fixplanets} tasks in CASA \texttt{v6.3}. We then performed four rounds of phase (\texttt{solint}=`inf', `120s', `60s', `30s') and one round of amplitude (\texttt{solint}=`inf') self-calibration after flagging channels containing strong lines, which resulted in a 6$\times$ improvement in the peak continuum SNR. We then applied the self-calibration tables to the line spectral windows. Finally, we subtracted the continuum using the \texttt{uvcontsub} task with a first-order polynomial.

For the ALMA data, we followed the same CLEANing approach as in the previous section. However, given the higher sensitivity and angular resolution of the observations, we instead selected a combination of \texttt{robust} value and \textit{uv}-taper that ensured a circularized beam of 0\farcs44, following the approach, e.g., of \citet{Czekala21}. This resulted in a line RMS noise of 1.7~mJy~beam$^{-1}$ in 0.22~km~s$^{-1}$ channels.

\begin{figure*}[]
\centering
\includegraphics[width=\linewidth]{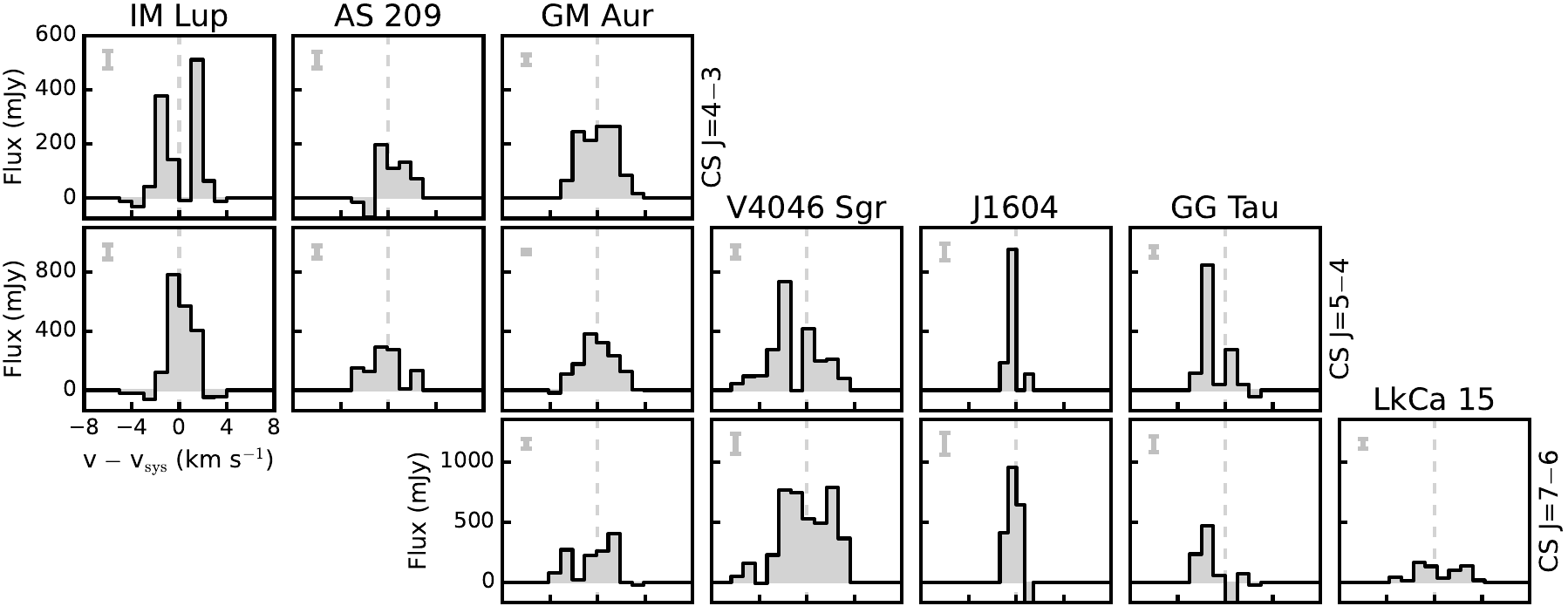}
\caption{Integrated intensity spectra of CS J=4--3 (\textit{top row}), J=5--4 (\textit{middle row}), and J=7--6 (\textit{bottom row}) obtained from our SMA observations. Each column corresponds to one disk. Spectra are extracted using a Keplerian mask based on the known disk geometry and vertical errorbars show the 1$\sigma$ RMS. The narrow linewidth of J1604 is due to the nearly face-on geometry of this disk.}
\label{fig:CS_fluxes_SMA}
\end{figure*}

\begin{figure*}[]
\centering
\includegraphics[width=\linewidth]{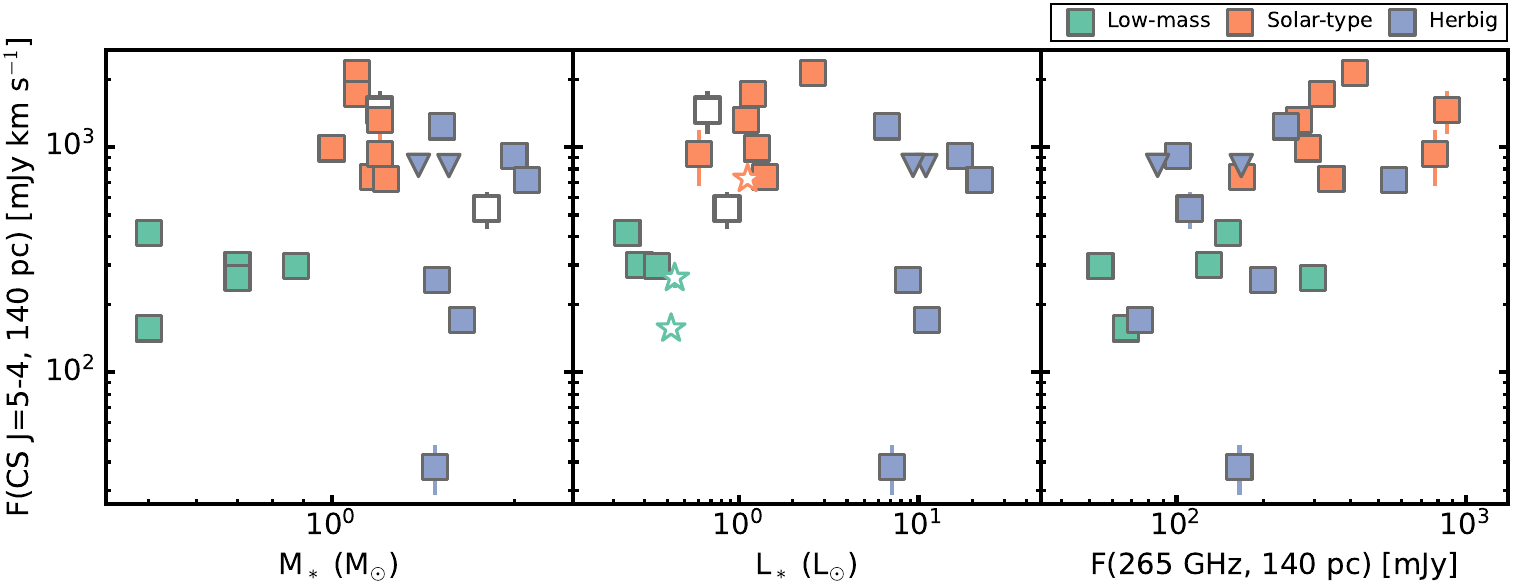}
\caption{CS J=5--4 integrated fluxes as a function of stellar mass (\textit{left}), stellar bolometric luminosity (\textit{middle}), and disk continuum flux at 265~GHz (\textit{right}). Green, orange, and blue colors mark low-mass T~Tauri (${<}1~M_{\odot}$), Solar-type ($1-1.3~M_{\odot}$), and Herbig (${>}1.3~M_{\odot}$) systems, respectively. The two gray square boxes correspond to V4046~Sgr and GG~Tau, which are known binary/multiple systems whose masses and luminosities represents the system totals. The star symbols indicate disks (DR~Tau, DG~Tau, DO~Tau) with known in-fall signatures and high accretion luminosities \citep[e.g.,][]{Manara14, McClure19}. Downward triangles indicate 3$\sigma$ flux upper limits. All fluxes have been scaled to a distance of 140~pc and continuum emission has been scaled to 265~GHz via F$_{\nu}\propto\nu^{2.2}$ \citep{Andrews20} for flux F$_{\nu}$ and frequency $\nu$. Fluxes are compiled from Table \ref{tab:obs-list} and additional literature sources \citep{Pietu11, LeGal19, Long19, Podio19, Podio20_DGTau, Facchini21PDS, Teague22_CS54_flux, Smirnov22, Pegues23, Law25}.}
\label{fig:CS_flux_comparison}
\end{figure*}

\subsection{ALMA and NOEMA Archival Data}
\label{sec:archival_data}

In addition to these new SMA and ALMA observations, we also compiled a variety of archival CS data, which we briefly describe on a line-by-line basis below. Full details are provided in Table \ref{tab:obs-list}.

\textit{CS J=2--1}: We acquired publicly-available data products, including disk-integrated fluxes, from the MAPS ALMA Large Program \citep{Oberg21_MAPSI, Czekala21, Legal21} for the IM~Lup, GM~Aur, AS~209, and MWC~480 disks. These data include ALMA Band 3 line observations at ${\approx}$0\farcs3 with 0.5~km~s$^{-1}$ channels. We also made use of published CS J=2--1 NOEMA data of the GG~Tau disk taken at ${\sim}$3-4$^{\prime \prime}$ with 0.19~km~s$^{-1}$ channels from \citet{Phuong21}.

\textit{CS J=3--2}: We obtained published CS J=3--2 data in the DM~Tau disk from \citet{Semenov18}, which were taken in ALMA Band 4 at ${\approx}$0\farcs5 and 0.30~km~s$^{-1}$.

\textit{CS J=5--4}: ALMA Band 6 observations of CS J=5--4 in the MWC~480 disk were taken from \citet{Legal21}, while those in the LkCa~15, DM~Tau, and CI~Tau disks are from \citet{LeGal19}. All data were at a comparable angular resolution of ${\approx}$0\farcs5-0\farcs6, but the MWC~480 data had a slightly higher velocity resolution of 0.18~km~s$^{-1}$ compared to the other disks at 0.5~km~s$^{-1}$.

\textit{CS J=6--5}: We obtained CS J=6--5 data of the MWC~480, LkCa~15, and V4046~Sgr disks from the ALMA Band 7 observations presented in \citet{LeGal19}. All three disks were observed at ${\approx}$0\farcs4-0\farcs7, but V4046~Sgr\footnote{The CS J=6--5 data of the V4046~Sgr disk is from ALMA program 2017.1.00938.S (PI: R. Loomis), which is incorrectly attributed in \citet{LeGal19}.} had a considerably higher velocity resolution of 0.1~km~s$^{-1}$ versus the 1~km~s$^{-1}$ channels of the MWC~480 and LkCa~15 disks.

\textit{CS J=7--6}: We took CS J=7--6 data of the CI~Tau disk from the ALMA Band 7 observations previously published in \citet{Rosotti21} at ${\approx}$0\farcs1 and 0.5~km~s$^{-1}$.

\section{Results} \label{sec:results}

\subsection{Computing Disk-Integrated Fluxes}  \label{sec:integrated_fluxes}

The primary aim of this work is to determine the disk-integrated CS rotational temperatures and column densities in our sample. To do so, we first need to measure the integrated CS line fluxes. 

For our SMA disk sample, we computed line fluxes from spectra extracted using the \texttt{gofish} tool \citep{Teague19JOSS} and the same Keplerian masks employed during CLEANing. The maximum radius of each mask is listed in Table \ref{tab:obs-list} along with the measured line fluxes. We used a bootstrapping approach to estimate uncertainties, which we took to be the standard deviation of the integrated fluxes within the same Keplerian masks but spanning only line-free channels in the image cubes. We applied 500 randomly-generated line-free masks to compute the flux uncertainty. Figure \ref{fig:CS_fluxes_SMA} shows a gallery of the CS spectra extracted from our SMA observations. Appendix \ref{sec:app:mom0_plots} presents a gallery of zeroth moment maps of all new CS transitions used in this work.

For the previously-published ALMA or NOEMA observations, we adopted the reported fluxes when available. We were careful to choose those fluxes measured within a consistent maximum radius, as indicated in Table \ref{tab:obs-list}. Several lines and sources did not have reported flux measurements (CS J=7--6 in CI~Tau) or only had fluxes measured over a portion of the full disk (CS J=5--4 in MWC~480 and CS J=6--5 in V4046~Sgr). For these cases, we instead consistently computed the total flux using the method described above.

Overall, we find a wide range in integrated CS line fluxes, from the relatively faint CS J=2--1 emission of ${\approx}$50~mJy~km~s$^{-1}$ in the MWC~480 disk to the bright CS J=7--6 flux of ${>}$1~Jy~km~s$^{-1}$ in the V4046~Sgr disk. In general, higher J transitions are stronger, by up to an order of magnitude, than lower J lines, consistent with their increasing line strengths. The inclusion of sensitive ALMA data allows for the detection of fainter, lower-J lines, while the SMA data provide broad spectral coverage of several CS lines simultaneously.

\begin{figure*}[]
\centering
\includegraphics[width=\linewidth]{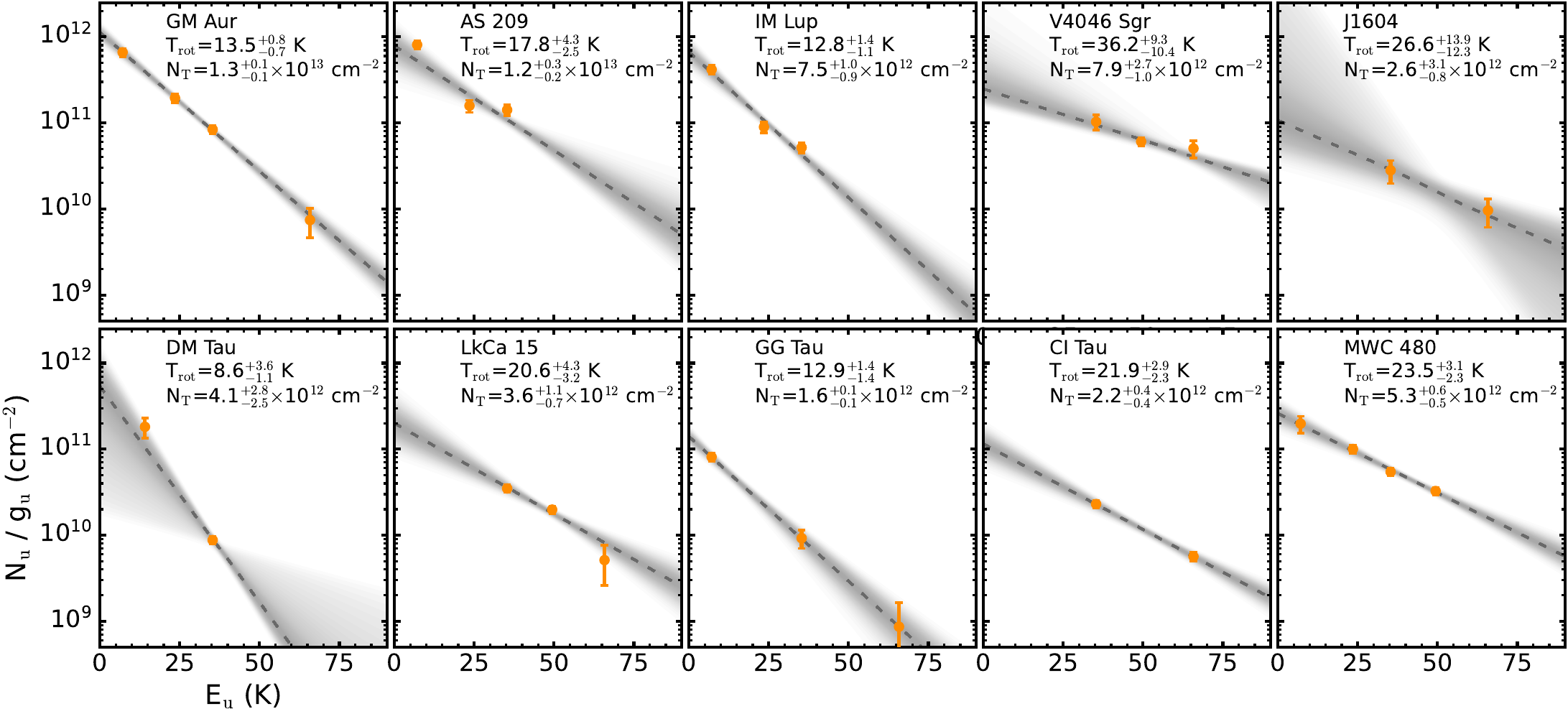}
\caption{Rotational diagrams of CS constructed using disk-integrated fluxes. Gray shaded regions show random draws from the fit posteriors. We included a 10\% calibration uncertainty on all measured line fluxes.}
\label{fig:rot_gallery}
\end{figure*}

\begin{figure*}[]
\centering
\includegraphics[width=.95\linewidth]{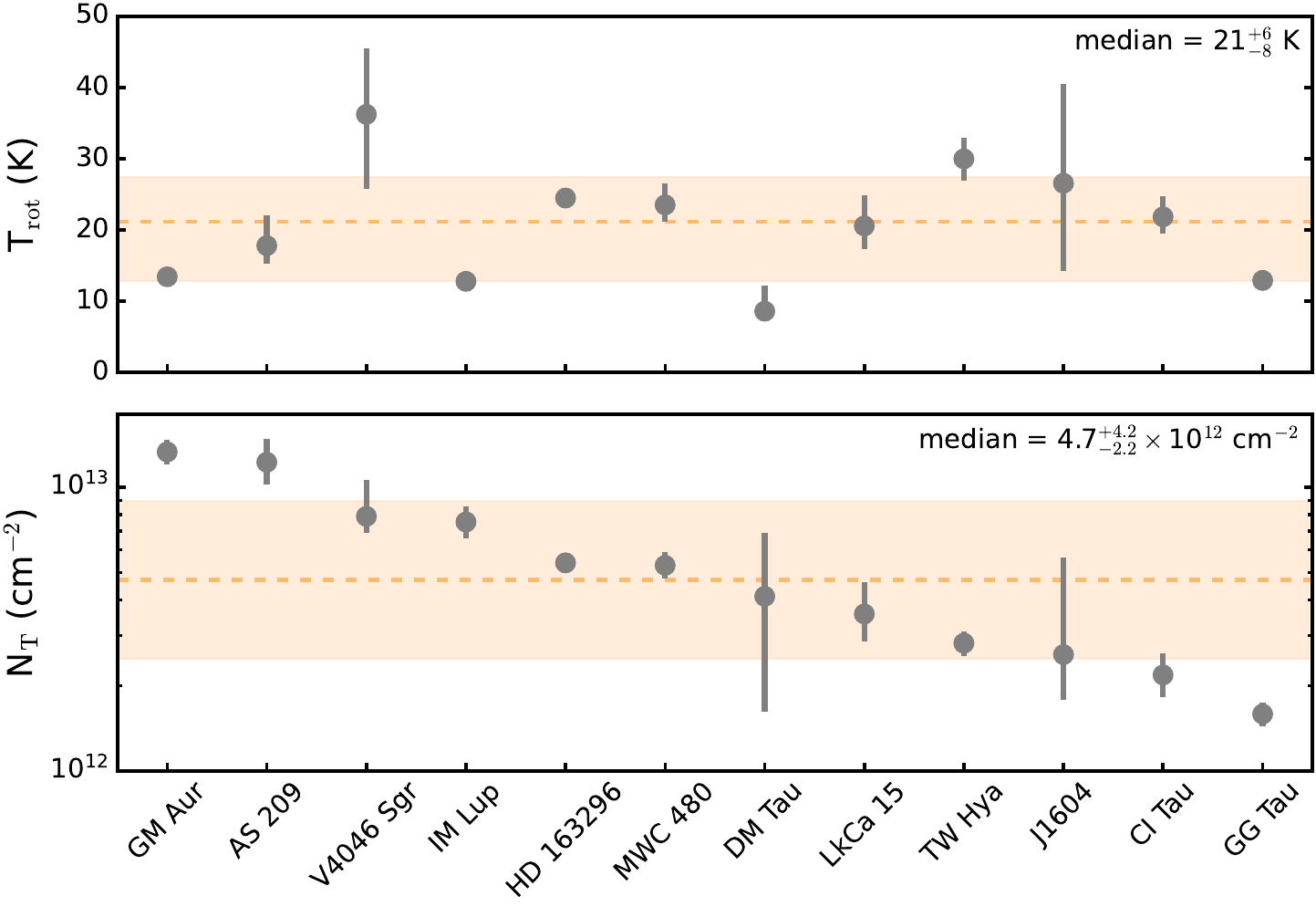}
\caption{Derived CS rotational temperatures (\textit{top}) and column densities (\textit{bottom}) for our disk sample, sorted from left to right by column density. The median values are shown by the dashed orange lines and the shaded regions indicate the 16th-84th percentiles. Several points have uncertainties smaller than the markers. The values for the TW~Hya and HD~163296 disks are taken from \citet{Teague18_TWHya} and \citet{Law25}, respectively.}
\label{fig:gallery_Trot_Ncol}
\end{figure*}

\subsection{Comparison of CS Line Fluxes} \label{sec:disk_Avg_fluxes}

Before determining CS excitation conditions on a per-disk basis, we searched for trends in the measured disk-integrated fluxes with stellar properties and the disk continuum flux, as the former may influence the CS flux due to the differing radiation environments and the latter serves as a proxy for the total disk mass. Here, we focused on the CS J=5--4 line, since it has been observed in all disks in our sample (see Table \ref{tab:obs-list}). The first two panels in Figure \ref{fig:CS_flux_comparison} show distance-normalized disk-integrated fluxes as a function of stellar host mass and bolometric luminosity. To augment our sample, we also included literature CS J=5--4 flux measurements of DO~Tau \citep{LeGal19}, DG~Tau \citep{Podio19}, DR~Tau \citep{Huang24}, TW~Hya \citep{Teague18_TWHya}, and PDS~70 \citep{Facchini21PDS}. We also included CS J=5--4 detections for the following Herbig disks: HD~163296 \citep{Law25}; HD~100453, HD~139614, HD~142666, and HD~145718 \citep{Smirnov22}; and upper limits for the HD~36112 and HD~144432 disks\footnote{The CS J=5--4 upper limit for the HD~34282 disk from \citet{Pegues23} is omitted as it is relatively unconstraining considering the far distance (306~pc) of this system. We also excluded the upper limit of HD~141569 from \citet{Smirnov22} due to this system hosting a hybrid disk at an evolutionary state between the protoplanetary and debris-disk regimes.} \citep{Pegues23}.

The CS J=5--4 flux generally increases with stellar mass and luminosity for low-mass (${<}1$~M$_{\odot}$) and Solar-mass (1-1.3~M$_{\odot}$) T~Tauri systems. However, disks around more massive Herbig stars (${>}1.3$~M$_{\odot}$) show  reduced fluxes, sometimes significantly, compared to Solar-mass systems and exhibit considerable variation, spanning an order-of-magnitude at a given mass or luminosity. Notably, the lowest CS J=5--4 flux is reported in the disk around the Herbig source HD~100453 \citep{Smirnov22}. Overall, the observed CS flux trends suggest three different regimes: fainter fluxes toward low-mass systems, increased fluxes in Solar-type disks, and highly-varied and often reduced fluxes toward Herbig disks.

Next, we also examined how CS fluxes vary as a function of millimeter continuum emission in the last panel in Figure \ref{fig:CS_flux_comparison}. CS fluxes generally increased with higher continuum fluxes, consistent with the findings of \citet{Pegues23}, who identified a similar positive trend in a smaller number of disks. This trend also shows relatively low scatter, i.e., a factor of a few, in flux, which is smaller than that of most other molecular lines reported in \citet{Pegues23}, which show one or more order-of-magnitude scatter in fluxes.

\subsection{Disk-Averaged Rotational Temperatures and Column Densities} \label{sec:Trot_Ncol_derivation}

To determine disk-averaged column densities and excitation temperatures, we used a rotational diagram analysis \citep{Goldsmith99}, following the approach of \citet{LeGal19, Legal21}. We assumed all lines are in local thermal equilibrium (LTE), as CS is expected to originate in the warm molecular layers of the disk where the gas densities are much higher than the CS line critical densities \citep[${\sim}$10$^4$-10$^6$~cm$^{-3}$;][]{Shirley15}. Then, we can relate the disk-integrated flux density, $S_{\nu} \Delta v$, measured across a solid angle, $\Omega$, to the column density of molecules in their respective upper energy state, N$_{\rm{u}}$ via:

\begin{equation}
    N_u = \dfrac{4 \pi S_{\nu} \Delta v}{A_{ul} \Omega hc},
\end{equation}

\noindent where A$_{ul}$ is the Einstein coefficient. Here, we use the disk-integrated flux density measured in a radius of R$_{\rm{max}}$, as listed in Table \ref{tab:obs-list}.

Then, the total column density N$_{\rm{T}}$ and rotational temperature T$_{\rm{rot}}$ are related to N$_u$ according to the Boltzmann equation:

\begin{equation}
    \dfrac{N_u}{g_u} = \dfrac{N_{\rm{T}}}{Q(T_{\rm{rot}})} e^{-E_u / k_{\rm{B}} T_{\rm{rot}}},
\end{equation}

\noindent where $g_u$ is the upper state degeneracy, $E_{\rm{u}}$ is the upper state energy, and Q is the molecular partition function. For a diatomic molecule like CS, we use the following partition function \citep[][]{Gordy84}:

\begin{equation}
Q(T_{\rm{rot}}) \approx \dfrac{k_B T_{\rm{rot}}}{h B_0} + \dfrac{1}{3}
\end{equation}

\noindent with a rotational constant B$_0=$ 24495.56 $\times 10^6$~Hz \citep{Muller05}. All spectroscopic line data were retrieved from the CDMS catalog, as listed in Table \ref{tab:obs-list}.

We then constructed a likelihood function $\mathcal{L}$(data, N$_{\rm{T}}$, T$_{\rm{rot}}$) based on $\chi^2$ statistics and sampled the posterior probability distributions for T$_{\rm{rot}}$ and N$_{\rm{T}}$ using the Markov Chain Monte Carlo (MCMC) code \texttt{emcee} \citep{Foreman_Mackey13}. All fits used thermal line widths of ${\approx}$0.2-0.3~km~s$^{-1}$ \citep[e.g.,][]{Teague18_TWHya, Paneque24}. In the fitting process, we also included optical depth correction factors, C$_{\tau}$ = $\tau / 1 - e^{-\tau}$, for the true level populations of each line. As C$_{\tau}$ can be written as a function of N$_u$, \citep[e.g.,][]{Loomis18}, this allows for a simultaneous estimate of the optical depths of individual transitions.

In general, we adopted broad uniform priors of T$_{\rm{rot}} (\rm{K}) = \mathcal{U}(3, 300)$ and 4-$\log_{10}$~N$_{\rm{T}} (\rm{cm}^{-2}) = \mathcal{U}(7, 20)$. For the DM~Tau, J1604, and CI~Tau disks, which only have two CS lines, we used restricted priors of (T$_{\rm{rot}} (\rm{K}) = \mathcal{U}(5, 50)$ and $\log_{10}$~N$_{\rm{T}} (\rm{cm}^{-2}) = \mathcal{U}(10, 14)$) to ease the fitting. All MCMC runs used 512 walkers with 2000 burn-in steps and an additional 1000 steps to sample the posterior distribution. The best-fit values and uncertainties are taken as the median, and 16th-84th percentiles, respectively, of the posterior samples. We also included a systematic flux calibration uncertainty of 10\% added in quadrature to the statistical uncertainty on all measured line fluxes.

Figure \ref{fig:rot_gallery} shows the fitted rotational diagrams for all disks. The majority of the CS lines are optically thin $(\tau < 0.2)$, with the exception of a few sources (IM~Lup, GM~Aur, AS~209, DM~Tau) where one or more lines reach $\tau \approx 0.5$-$0.8$. Of those disks in our sample, only two (MWC~480, LkCa~15) had existing, multi-line estimates of both T$_{\rm{rot}}$ and N$_{\rm{T}}$ \citep{LeGal19, Legal21}, while several others estimated N$_{\rm{T}}$ from a single CS transition and an assumed excitation temperature \citep{Semenov18, Phuong21}. Overall, our derived T$_{\rm{rot}}$ and N$_{\rm{T}}$ values are generally consistent with these previous measurements but have considerably smaller uncertainties due to the inclusion of additional CS transitions and the ability to constrain excitation temperatures.

Figure \ref{fig:gallery_Trot_Ncol} shows the derived T$_{\rm{rot}}$ and N$_{\rm{T}}$ for the entire disk sample, sorted by column density. To further augment our sample, we included the disk-averaged CS T$_{\rm{rot}}$ and N$_{\rm{T}}$ values derived from similar multi-line CS observations in the TW~Hya \citep{Teague18_TWHya} and HD~163296 disks \citep{Law25}. Since \citet{Teague18_TWHya} do not report disk-averaged values, we instead adopted the T$_{\rm{rot}}$ and N$_{\rm{T}}$ at a radius of 100~au. Across our sample, T$_{\rm{rot}}$ ranges from ${\approx}$10~K to 40~K with a median and 16th-84th percentile range of 21$^{+6}_{-8}$~K, while N$_{\rm{T}}$ spans approximately an order of magnitude from 10$^{12}$~cm$^{-2}$ to 10$^{13}$~cm$^{-2}$ with a median and 16th-84th percentile range of 4.7$^{+4.2}_{-2.2}\times$10$^{12}$~cm$^{-2}$.

\begin{figure*}[]
\centering
\includegraphics[width=\linewidth]{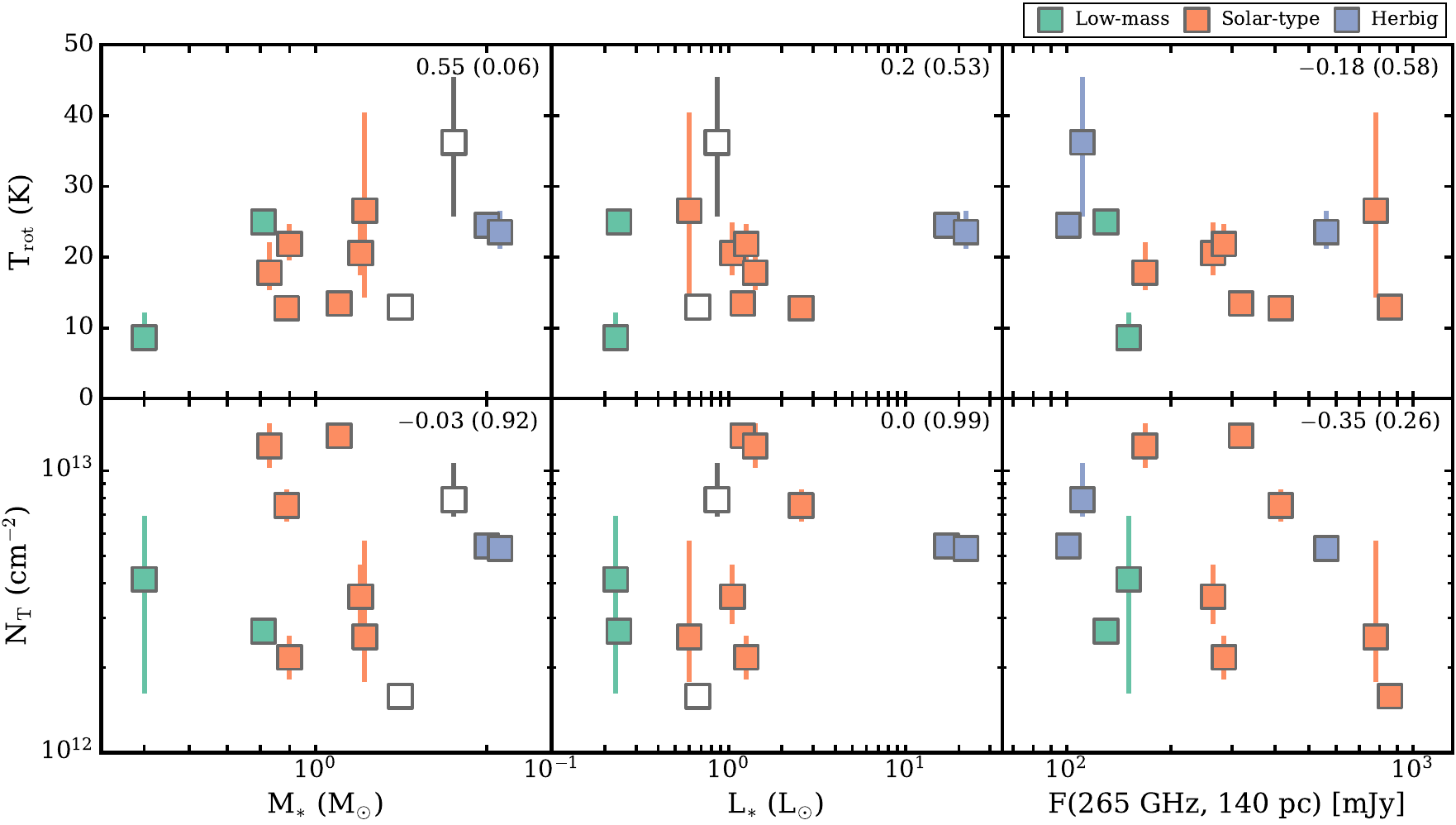}
\caption{CS rotational temperatures (\textit{top row}) and column densities (\textit{bottom row}) versus stellar mass (\textit{left column}), stellar bolometric luminosity (\textit{middle column}), and disk continuum flux at 265~GHz normalized to 140~pc (\textit{right column}). The two gray square boxes correspond to V4046~Sgr and GG~Tau, which are known binary/multiple systems whose masses and luminosities represents the system totals. Colors and symbols are as in Figure \ref{fig:CS_flux_comparison}. Pearson's correlation coefficients and associated p-values are shown in the upper right corner of each panel.}
\label{fig:correlation_Trot}
\end{figure*}

\section{Discussion} \label{sec:discussion}

Next, we use our disk sample to explore potential correlations with stellar and disk properties and in doing so, shed light on the chemical origins of CS in protoplanetary disks. First, we briefly summarize the expected formation and destruction mechanisms of CS, which will help inform later interpretations of observed sample trends.

\subsection{The Chemical Origins of CS} \label{sec:astrochemical_origins}

In the outer disk regions probed by our observations, CS chemistry is driven by the presence of ionized sulfur and gas-phase carbon \citep{Semenov18, LeGal19}. In detail, CS is expected to form in the disk upper layers via rapid ion-neutral, gas-phase reactions between S$^+$ and small hydrocarbons CH$_{\rm{x}}$ and C$_{\rm{y}}$H (with x=1-4 and y=2-3). These reactions, in turn, generate carbonated S-ions (C$_2$S$^+$, C$_3$S$^+$, HCS$^+$, H$_2$CS$^+$, H$_3$CS$^+$), which then dissociatively recombine with electrons to produce CS and other small neutral sulfuretted molecules. Slower formation routes also occur via barrierless neutral-neutral reactions involving atomic sulfur, including S + CH, CH$_2$, or C$_2$, which is expected to produce a secondary CS reservoir closer to the disk midplane. The primary destruction pathways of CS are photodissociation processes primarily in the disk atmosphere, protonation and charge transfer reactions (with H$^+$, H$_3^+$, HCO$^+$, He$^+$), and freeze-out onto grain surfaces at gas temperatures of $\lesssim$30~K.

\subsection{Exploring Sample-Wide Trends} \label{sec:correlations}

Given our large sample with homogeneously-derived, disk-averaged CS rotational temperatures and column densities, we aim to determine how source properties influence the observed disk sulfur chemistry. We consider the potential influence of stellar mass and bolometric luminosity as well as disk mass, size, and structure, including dust morphology and the presence of inner cavities in transition disks. We also examine several additional physical properties, such as the gas-phase carbon-to-oxygen (C/O) ratio and stellar X-ray luminosity, that have been shown to influence CS chemistry in disk chemical modeling \citep[e.g.,][]{Semenov18}. Below, we discuss each of these parameters in detail. To assess the importance of potential correlations, we also computed the Pearson's correlation coefficient using the \texttt{pearsonr} function in \texttt{scipy.stats} \citep{Virtanen_etal_2020}, which we summarize in Table \ref{tab:correlations_table}. \\

\noindent \textit{Stellar Properties:} Here, we first consider the potential impact of stellar properties on the observed CS chemistry, as the radiation environment varies substantially from low-mass to Herbig stars and may influence the relative importance of different CS formation and destruction pathways. The first two columns of Figure \ref{fig:correlation_Trot} show CS T$_{\rm{rot}}$ and N$_{\rm{T}}$ as a function of stellar mass and bolometric luminosity. Previously, \citet{Legal21} compiled a disk sample with CS column densities by computing N$_{\rm{T}}$ over a range of assumed excitation temperatures (from 10-30~K) due to the lack of multi-line CS data and did not find any obvious trends with stellar mass or spectral type.

Here, we derived both T$_{\rm{rot}}$ and N$_{\rm{T}}$ for all disks in our sample, and likewise, identify no clear trends between N$_{\rm{T}}$ and either the stellar mass or bolometric luminosity. Similar to what was seen for CS fluxes (Figure \ref{fig:CS_fluxes_SMA}), Solar-mass systems have highly varied N$_{\rm{T}}$ values with almost an order-of-magnitude scatter. Although our sample only contains two disks around Herbig stars (HD~163296, MWC~480), both have comparable or lower N$_{\rm{T}}$ values than at least half of the Solar-mass systems. This may suggest that the stronger UV fields around Herbig stars lead to significant photodissociation -- a known destruction pathway of CS -- which reduces the total CS column density. Alternatively, this may be related to the environments around Herbig stars, which have high enough gas temperatures to substantially inhibit ice chemistry, resulting in lower CO depletion levels compared to T~Tauri disks \citep[e.g.,][]{Miotello23,Trapman25}. One potential consequence of this would be the reduction of available gas-phase carbon needed to form CS, as more elemental carbon is bound up in CO. We do, in fact, identify a tentative positive trend between CS T$_{\rm{rot}}$ and stellar mass, which likely reflects these warmer gas temperatures around higher-mass stars. However, given that CS column densities do not vary by more than an order of magnitude across all stellar spectral types in our sample, these effects are not likely to be dominant in setting the overall CS chemistry of Herbig sources.

Interestingly, if we consider only the stellar X-ray luminosity rather than the bolometric luminosity, we find a positive L$_{\rm{X}}$--N$_{\rm{T}}$ correlation across our sample, which implies that X-ray irradiation levels play a key role in CS formation. We discuss the implications of this correlation in detail in Section \ref{sec:Xray_CS}.\\

\begin{figure}[]
\centering
\includegraphics[width=\linewidth]{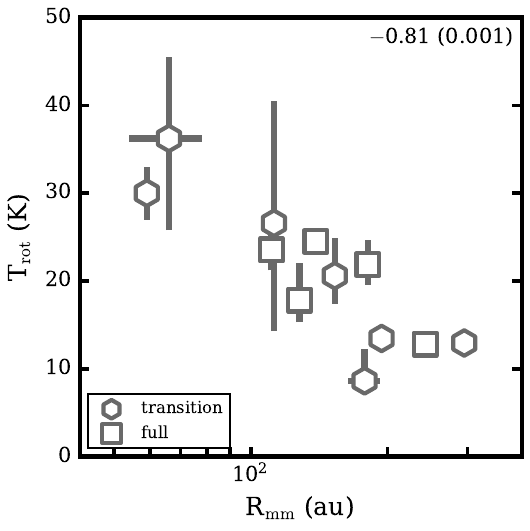}
\caption{CS rotational temperatures versus the millimeter dust disk sizes complied from \citet{Andrews14, Long18, Law21_MAPSIII, Long22, Curone25}. Disk sizes represent the radius containing 90\% of total flux. Transition disks are marked by hexagons. The Pearson's correlation coefficient and associated p-value are shown in the upper right.}
\label{fig:correlation_size}
\end{figure}

\noindent \textit{Disk Mass, Size, and Structure:} The millimeter dust content and structure of disks is expected to influence the underlying chemistry by altering, e.g., the gas temperature distribution and irradiation levels \citep[][]{Cleeves16, Facchini17}. The last column of Figure \ref{fig:correlation_Trot} shows CS T$_{\rm{rot}}$ and N$_{\rm{T}}$ versus the distance-normalized millimeter continuum flux. In our sample, continuum fluxes span more than an order of magnitude and in general, provide an approximate measure of the total disk mass \citep{Beckwith90}. Neither T$_{\rm{rot}}$ or N$_{\rm{T}}$ shows any dependence on the continuum flux, indicating that CS abundances do not necessarily trace the bulk disk mass but rather are set by other chemical and physical processes.

Broadly, the CS column density also does not strongly depend on the underlying disk structure either. We find no significant correlations between CS N$_{\rm{T}}$ and either the size of the millimeter dust disk or the radial extent of the inner dust cavities among the transition disks in our sample (see Appendix Figure \ref{fig:correlation_cavity_size}). Likewise, there are no systematic difference in either the T$_{\rm{rot}}$ or N$_{\rm{T}}$ in transition disks versus full, non-transition disks.

As shown in Figure \ref{fig:correlation_size}, the CS rotational temperatures do, however, show a negative correlation with the millimeter disk size. This suggests active CS chemistry even at large disk radii and at cooler gas temperatures. When combined with the lack of N$_{\rm{T}}$ correlation with disk size (Appendix Figure \ref{fig:correlation_Rmm_NT}), this implies that CS freeze-out may not be a dominant destruction mechanism. This is also consistent with the AS~209 and GM~Aur disks having the two highest disk-integrated N$_{\rm{T}}$ but among the coolest T$_{\rm{rot}}$ of ${\approx}$14-18~K. When combined with the fact that CS chemistry does not show strong trends with stellar spectral types, as discussed above, this correlation further supports the idea that CS chemistry is not highly sensitive to disk gas temperatures.

\begin{deluxetable}{lccc}
\tablecaption{Correlations in our Disk Sample \label{tab:correlations_table}}
\tablewidth{0pt}
\tablehead{
\colhead{} & \colhead{T$_{\rm{rot}}$ (CS)}  & \colhead{N$_{\rm{T}}$ (CS)}
}
\startdata
M$_*$                              & \textit{0.55 (0.06)} & \ldots \\
L$_*$                              & \ldots               & \ldots \\
L$_{\rm{X}}$                       & \ldots               & 0.84 (1.3$\times$10$^{-3}$) \\
F$_{\rm{mm}}$                      & \ldots               & \ldots \\
R$_{\rm{mm}}$                      & $-$0.81 (0.001)       & \ldots \\
R$_{\rm{cavity}}$\tablenotemark{a} & \ldots                & \ldots \\
C/O ratio\tablenotemark{b}         &  \ldots              & \ldots \\
\enddata
\tablecomments{Pearson's correlation coefficients with p-values in parenthesis are only indicated for significant (p$<$0.05) correlations. Those between 5\% and 10\% significance are considered tentative and indicated in italics.}
\tablenotetext{a}{No correlations were found within the transition disk sample alone or if we considered the full disks to have R$_{\rm{cavity}}=0$.}
\tablenotetext{b}{As noted in the text, only six disks in our sample have well-constrained gas-phase C/O ratios.}
\end{deluxetable}

\noindent \textit{C/O Ratios:} Given the role that carbon plays in the formation of CS, the gas-phase C/O ratio is an important property to consider \citep[e.g.,][]{Semenov18, LeGal19,Williams25}. In particular, increases in C/O are predicted to result in higher CS column densities, with the CS/SO ratio proposed as an empirical tracer of the C/O ratio \citep[e.g.,][]{Legal21, RMarichalar22}.

Six disks in our sample have detailed estimates of their elemental C/O ratios derived from sensitive molecular line observations. These measurements fall into two broad categories of C/O${\approx}$1.5-2.0 \citep[AS~209, HD~163296, MWC~480, TW~Hya;][]{Bergin16, Bosman21_CO, Alarcon21} and C/O${\approx}$0.8-1.0 \citep[IM~Lup, LkCa~15;][]{Cleeves18, Sturm22}. Comparing these two groupings, we find no notable differences in either the CS T$_{\rm{rot}}$ or N$_{\rm{T}}$ (see Appendix Figure \ref{fig:correlation_CO_Ratio}). The lack of a clear trend could be due to a variety of compounding effects, including the degree of sulfur depletion (i.e., S/H ratio), uncertainties in the derived C/O ratios, or the small range in C/O ratios covered in our sample, which consists of only large and massive disks.

\begin{figure*}[]
\centering
\includegraphics[width=\linewidth]{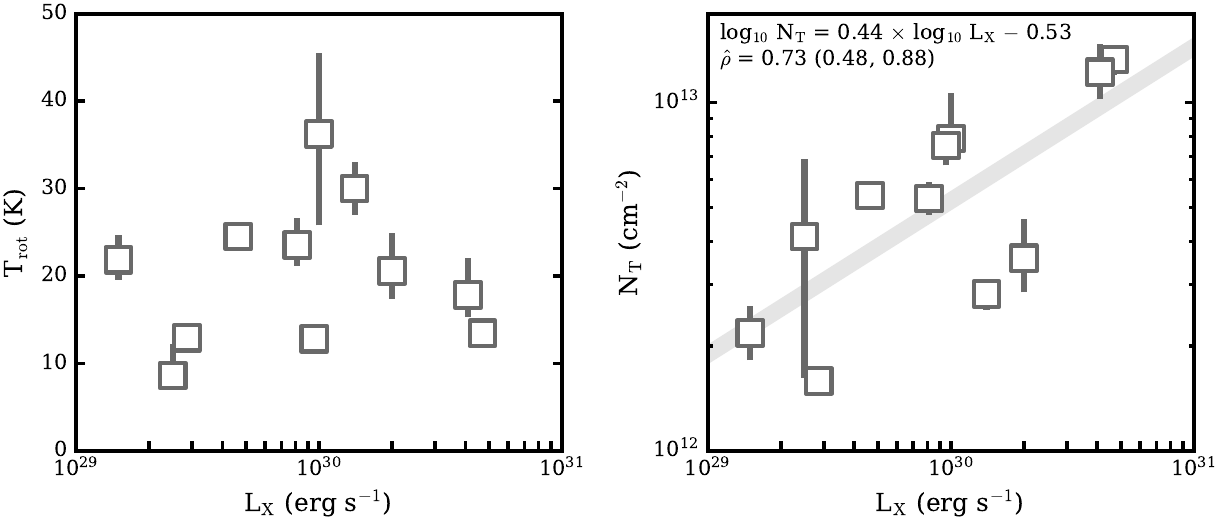}
\caption{CS rotational temperatures (\textit{left}) and column densities (\textit{right}) versus the stellar X-ray luminosity. For the N$_{\rm{T}}$-L$_{\rm{X}}$ panel, the best-fit relation is marked with a solid grey line.}
\label{fig:correlation_Xray}
\end{figure*}

\subsection{X-ray Driven CS Chemistry} \label{sec:Xray_CS}

Differences in stellar X-ray luminosity are predicted to influence the sulfur chemistry in disks \citep[e.g.,][]{Semenov18}. The stellar hosts in our sample span L$_{\rm{X}}$ from ${\approx}$10$^{29}$~erg~s$^{-1}$ to ${\approx}$5$\times$10$^{30}$~erg~s$^{-1}$ with existing L$_{\rm{X}}$ measurements for all sources except J1604. Here, we aim to investigate how the CS gas conditions respond to changes in the levels of X-ray irradiation. 

In Figure \ref{fig:correlation_Xray}, we compare our derived CS T$_{\rm{rot}}$ and N$_{\rm{T}}$ values versus the stellar X-ray luminosity. While no trend is observed for T$_{\rm{rot}}$, we identify a positive correlation between the CS column density and L$_{\rm{X}}$. To better quantify this L$_{\rm{X}}$-N$_{\rm{T}}$ correlation, we employed the Bayesian linear regression method of \citet{Kelly07} using the \texttt{linmix} python implementation\footnote{\url{https://github.com/jmeyers314/linmix}}. We find a best-fit relation of $\log_{10}$ N$_{\rm{T}} = (0.44\pm0.17$) $\times$ $\log_{10}$~L$_{\rm{X}} + (-0.53 \pm 5.1)$ with a correlation coefficient of $\hat{\rho}=0.73$ and associated confidence intervals of (0.48, 0.88), which represent the median and 16-84th confidence regions, respectively, of the 2.5$\times$10$^6$ posterior samples for the regression. We note that the mean correlation coefficient is different than that reported in Table \ref{tab:correlations_table}, as this regression accounts for uncertainties for a more robust quantification of this correlation. Figure \ref{fig:correlation_Xray} shows the derived relationship.

Surprisingly, the only chemical models \citep{Semenov18} to directly explore the role of X-rays on disk CS content find the opposite trend between CS column density and L$_{\rm{X}}$. By decreasing L$_{\rm{X}}$ from 10$^{31}$ to 10$^{29}$~erg~s$^{-1}$ -- the same range covered in our sample -- \citet{Semenov18} predict nearly an order-of-magnitude increase in CS column density. It is not clear what reactions are ultimately responsible for this change in the models of \citet{Semenov18}, but we stress that this is the opposite of what we observe, i.e., increased L$_{\rm{X}}$ corresponds to enhanced CS column densities. Instead, we hypothesize that higher levels of X-ray irradiation increase production of He$^+$ ions \citep[e.g.,][]{Cleeves15_Xray}, which, in turn, destroy gas-phase CO and liberate extra carbon to participate in CS-formation chemistry. X-rays can also boost the abundances of ions, such as S$^+$, CS$^+$, and HCS$^+$, that are involved in CS production \citep[e.g.,][]{Glassgold97, Aikawa99, Henning10, Teague15_Xray, Waggoner22}. Simulations of chemistry driven by X-ray flares found that sulfur-bearing molecules were indeed susceptible to X-ray flaring events, with increased abundance of organosulphur species in those models with a history of X-ray flares \citep{Waggoner22}. Thus, the existence of a positive L$_{\rm{X}}$-N$_{\rm{T}}$ correlation suggests that the primary CS formation route is via ion-neutral reactions in the upper disk layers, where X-ray-induced enrichment of S$^+$ and C$^+$ drives abundant CS production. While additional dedicated models focused on X-ray-driven CS chemistry are needed to fully explore this relationship, this correlation suggests that the CS column density traces stellar X-ray luminosity in young disk systems.

\begin{figure*}[]
\centering
\includegraphics[width=\linewidth]{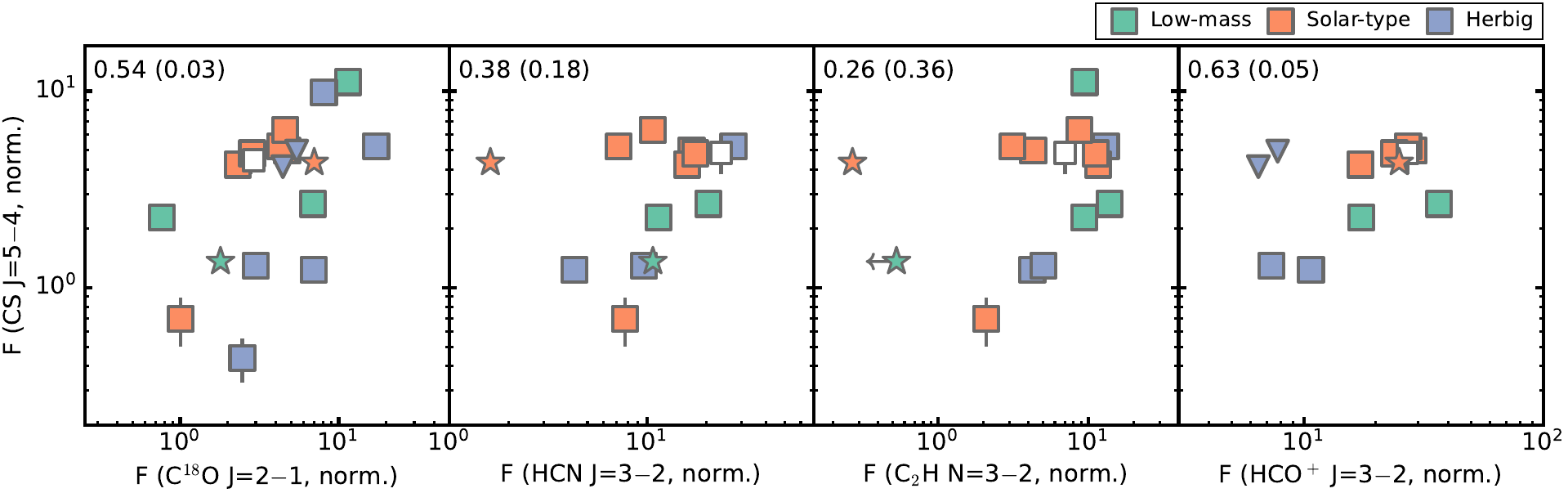}
\caption{CS J=5--4 correlations plots with C$^{18}$O J=2--1, HCN J=3--2, C$_2$H N=3--2, and HCO$^+$ J=3--2. All fluxes are normalized by the disk-integrated continuum flux. Pearson's correlation coefficients and corresponding p-values are computed excluding upper limits and marked in the upper left of each panel. The stars indicate the DR~Tau and DO~Tau disks, which have known in-fall signatures and high accretion luminosities. Colors and symbols are as in Figure \ref{fig:CS_flux_comparison}. Fluxes are compiled from Table \ref{tab:obs-list} and additional literature sources \citep{Pietu11, LeGal19, Long19, Podio19, Podio20_DGTau, Facchini21PDS, Teague22_CS54_flux, Smirnov22, Pegues23, Huang23, Huang24, Stapper24, Law25}.} \label{fig:correlation_lines_flux}
\end{figure*}
\subsection{How Does CS Compare with Other Molecular Tracers?}

Our sample is composed of disks whose chemical composition has been well-characterized in a variety of oxygen-, nitrogen-, and carbon-bearing species. Here, we explore possible chemical connections with the CS chemistry by considering the commonly-observed molecular tracers C$^{18}$O, HCN, C$_2$H, and HCO$^+$. As robust column density estimates in these additional tracers only exist for a subset of disks in our sample, we instead opted to explore correlations between line fluxes. Following \citet{Bergner19}, we normalized the disk-integrated line fluxes by the continuum flux to limit possible dependence on the underlying disk mass. As in Section \ref{sec:disk_Avg_fluxes}, we chose CS J=5--4 for this comparison, since this line is observed in all disks in our sample and approximately matches the excitation properties of the other tracers. We also included the same literature sources as in Section \ref{sec:disk_Avg_fluxes} in this comparison, but note that not all disks have existing flux measurements for all lines of interest.

Figure \ref{fig:correlation_lines_flux} shows each of the normalized line fluxes plotted against that of CS. We find a tentative positive correlation with both C$^{18}$O and HCO$^+$, while no significant correlation is evident in C$_2$H and HCN. The CS-C$^{18}$O trend is likely driven by the fact that a larger C$^{18}$O reservoir provides more raw material to form CS, e.g., via X-ray produced He$^+$ ions that destroy CO molecules and generate extra gas-phase carbon. In contrast, HCO$^+$ is thought to be involved in a potential destruction route of CS, but if this route was dominant, then we might have expected to see a negative correlation. Instead, we see a modest positive trend, which may be explained by the dependence of HCO$^+$ on X-ray ionization and CO abundance \citep{Aikawa01}, which we find to both positively correlate with CS. Although not significant, there is a modest but highly-scattered positive trend in both C$_2$H and HCN. If additional data were to bear out either of these trends, they could naturally be explained due to known chemical links \citep[e.g.,][]{Bergner18}. For instance, C$_2$H is one of several small hydrocarbons that can act as reactants to produce both CS and HCN. Moreover, CS and HCN may be indirectly linked due to, e.g., the stellar X-ray luminosity, as one possible HCN formation pathway may involve X-ray photodesorption of CH$_3$CN-rich ices \citep{Basalg23}.

An important caveat to this analysis is that line fluxes may not always provide reliable tracers of molecular abundances due to, e.g., line optical depths or chemical effects. Ultimately, comprehensive disk samples with disk-integrated column densities, and abundances with respect to H$_2$, will be required to robustly establish or reject the existence of chemical relations between these species and CS.

\section{Conclusions} \label{sec:conlcusions}

We present a multi-line analysis of CS gas excitation in a sample of 12 protoplanetary disks compiled from new and archival observations obtained with the SMA, NOEMA, and ALMA. We conclude the following: 

\begin{enumerate}
    \item We present SMA observations of 14 new CS line detections, including CS J=4--3, J=5--4, J=7--6, across 7 disks and an ALMA detection of CS J=4--3 in the MWC~480 disk. When combined with existing data, we assembled a 12-disk sample with most sources having between 2-4 lines that span several 10s of K in E$_{\rm{u}}$.  
    \item We find that CS fluxes increase with stellar mass and bolometric luminosity for low-mass (${<}$1~M$_{\odot}$) and Solar-mass (1-1.3~M$_{\odot}$) T~Tauri systems, while disks around more massive (${>}$1.3~M$_{\odot}$) Herbig stars have highly-varied and often reduced fluxes.
    \item We derive CS rotational temperatures and column densities for our disk sample. Rotational temperatures span ${\approx}$10-40~K with a median of $21^{+6}_{-8}$~K, while column densities are between 10$^{12}$-10$^{13}$~cm$^{-2}$ with a median of $4.7^{+4.2}_{-2.2}\times10^{12}$~cm$^{-2}$.
    \item CS column densities do not show strong trends with most source properties, which broadly indicates that CS chemistry is not highly sensitive to disk structure or stellar spectral types.
    \item The positive correlation between CS column density and stellar X-ray luminosity highlights the critical role of ionization-driven chemistry via X-ray-enhanced S$^+$ and C$^+$ in setting the sulfur inventory of planet-forming disks.
    \item Tentative correlations between CS-C$^{18}$O and CS-HCO$^+$ line fluxes further support CS formation routes that benefit from gas-phase carbon librated from CO destruction via X-ray-produced He$^+$ ions.
    \end{enumerate}

Our findings suggest that stellar X-ray luminosity plays a decisive role in driving CS production in planet-forming disks, highlighting that ionization-driven chemical processes -- rather than purely thermal or structural disk properties -- are central to shaping the volatile sulfur reservoir during the epoch of planet formation. This study also demonstrates the necessity of multi-line CS observations to robustly constrain the gas-phase sulfur chemistry in protoplanetary disks. Since gas conditions are known to vary on small-scales across disks, an important next step will be spatially-resolved observations to discern how CS chemistry responds to local physical and chemical variations, e.g., due to embedded young protoplanets. In parallel to this, observations of more diverse disks are required. While Solar-mass systems were well-represented in our sample, relatively few disks around both low-mass and Herbig stars have had their sulfur chemistry surveyed in detail. In the near future, the ongoing DECO (2022.1.00875.L), CHEER (2024.1.01001.L), and DiskStrat (2024.1.01212.L) ALMA Large Programs will vastly increase disk demographics in both of these stellar-mass regimes, respectively.

As sulfur chemistry influences both prebiotic chemistry and planetary atmospheres, understanding these processes offers crucial insights into the chemical evolution and potential habitability of emerging planetary systems. Future multi-wavelength studies that combine X-ray, IR, and sub-mm observations will also be key to fully unraveling how high-energy radiation shapes disk chemistry across different disk environments.

\begin{acknowledgments}
The authors thank the anonymous referee for valuable comments that improved both the content and presentation of this work. This paper makes use of the following ALMA data: ADS/JAO.ALMA\#2013.1.01070.S, 2015.1.00296.S, 2016.1.00627.S, 2017.A.00014.S, 2017.1.00938.S, 2018.1.01055.L, and 2021.1.00899.S. ALMA is a partnership of ESO (representing its member states), NSF (USA) and NINS (Japan), together with NRC (Canada), MOST and ASIAA (Taiwan), and KASI (Republic of Korea), in cooperation with the Republic of Chile. The Joint ALMA Observatory is operated by ESO, AUI/NRAO and NAOJ. The National Radio Astronomy Observatory is a facility of the National Science Foundation operated under cooperative agreement by Associated Universities, Inc. Support for C.J.L. and F.L. was provided by NASA through the NASA Hubble Fellowship grant Nos. HST-HF2-51535.001-A and HST-HF2-51512.001-A, respectively, awarded by the Space Telescope Science Institute, which is operated by the Association of Universities for Research in Astronomy, Inc., for NASA, under contract NAS5-26555. 

This paper also makes use of data obtained with the SMA. The authors wish to recognize and acknowledge the very significant cultural role and reverence that the summit of Maunakea has always had within the indigenous Hawaiian community. We are most fortunate to have had the opportunity to conduct observation from this mountain.

We thank John Ilee and Giovanni Rosotti for sharing ALMA image cubes of the CI~Tau disk and Dmitry Semenov for sharing data of the DM~Tau disk.

\end{acknowledgments}

%

\facilities{ALMA, SMA, NOEMA}


\software{Astropy \citep{astropy_2013,astropy_2018}, \texttt{bettermoments} \citep{Teague18_bettermoments}, CASA \citep{McMullin_etal_2007, CASA22}, \texttt{GoFish} \citep{Teague19JOSS}, \texttt{keplerian\_mask} \citep{rich_teague_2020_4321137}, \texttt{linmix} \citep{Kelly07}, \texttt{pyuvdata} \citep{Hazelton17}, Matplotlib \citep{Hunter07}, MIR (\url{https://lweb.cfa.harvard.edu/~cqi/mircook.html}), NumPy \citep{vanderWalt_etal_2011}, \texttt{scipy} \citep{Virtanen_etal_2020}}



\clearpage
\appendix

\section{Gallery of Zeroth Moment Maps} \label{sec:app:mom0_plots}

\begin{figure*}[b]
\centering
\includegraphics[width=0.62\linewidth]{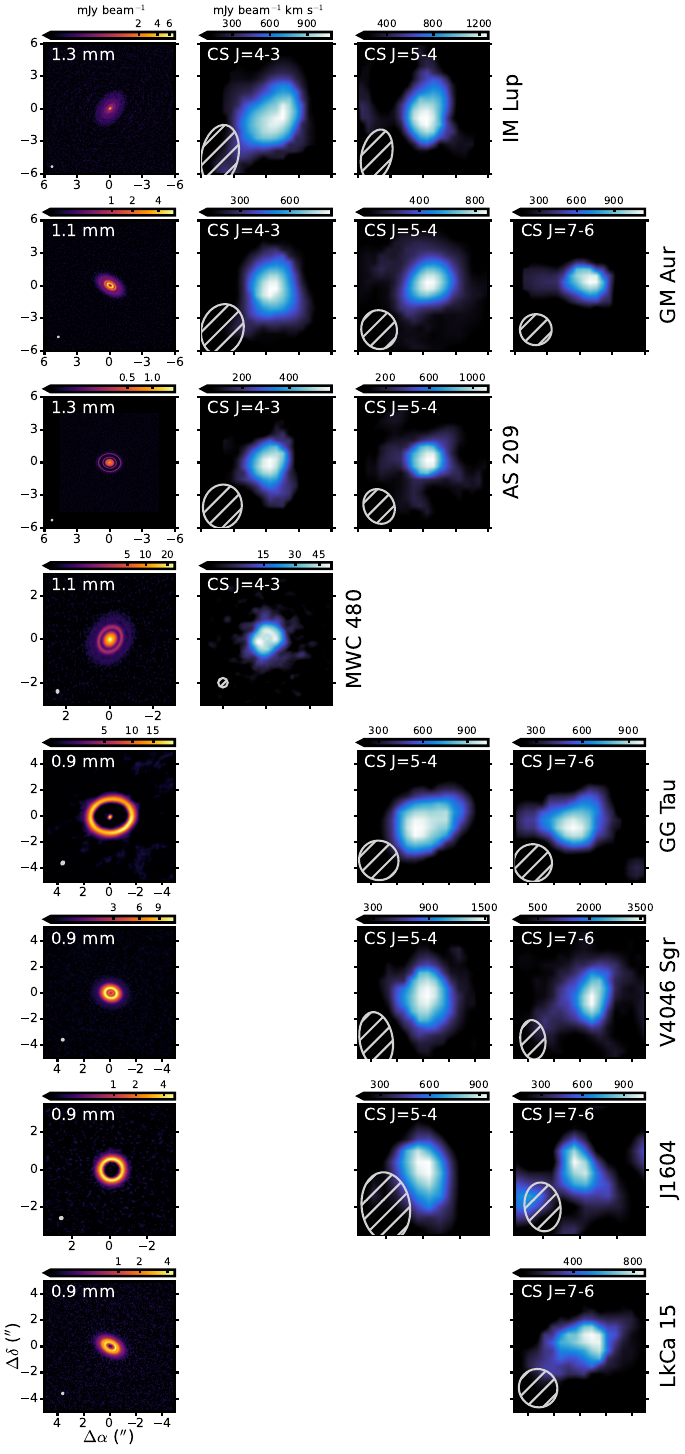}
\vspace{-10pt}
\caption{Gallery of continuum images \citep{Sierra21,Tang23,Curone25} and zeroth moment maps of all new CS lines presented in this work. The synthesized beam is shown in the lower left corner of each panel.}
\label{fig:mom0_app_plot}
\end{figure*}

Figure \ref{fig:mom0_app_plot} shows a gallery of velocity-integrated intensity, or ``zeroth moment", maps for all new CS line observations presented in this work. We generated all zeroth moment maps with the \texttt{bettermoments} \citep{Teague18_bettermoments} code using Keplerian masks.

\section{Additional Correlation Plots} \label{sec:app:corr_plots}

\begin{figure*}[b]
\centering
\includegraphics[width=\linewidth]{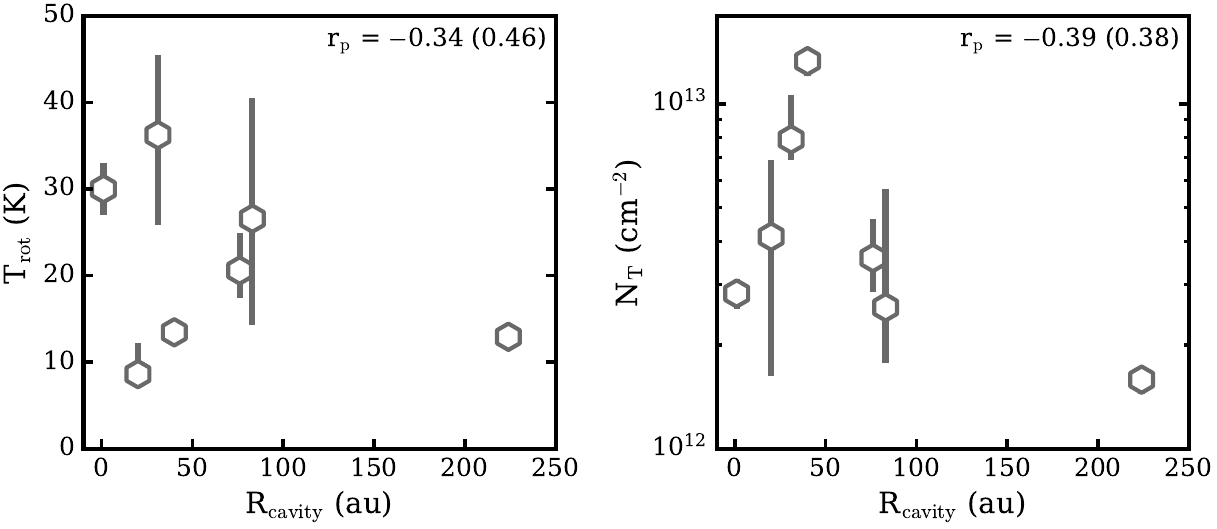}
\caption{CS rotational temperatures (\textit{left}) and column densities (\textit{right}) versus the millimeter cavity size of transition disks. Pearson's correlation coefficients and corresponding p-values are marked in the upper right of each panel.}
\label{fig:correlation_cavity_size}
\end{figure*}

\begin{figure*}[b]
\centering
\includegraphics[width=\linewidth]{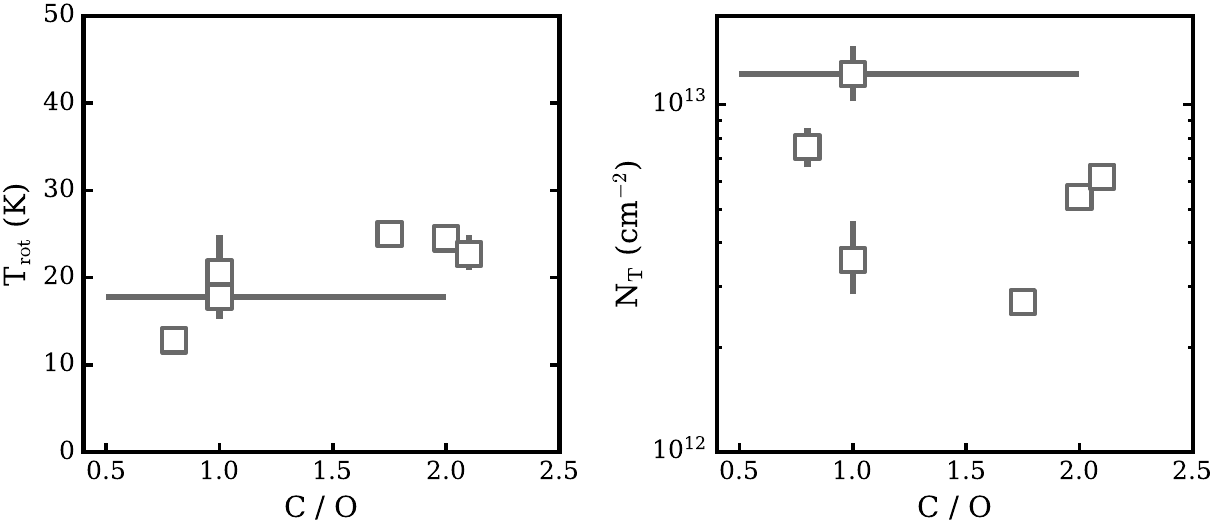}
\caption{CS rotational temperatures (\textit{left}) and column densities (\textit{right}) versus the gas-phase C/O ratio. The AS~209 disk has a radially-varying C/O ratio \citep{Alarcon21}, which is indicated as a horizontal line.}
\label{fig:correlation_CO_Ratio}
\end{figure*}

\begin{figure}[b]
\centering
\includegraphics[width=.5\linewidth]{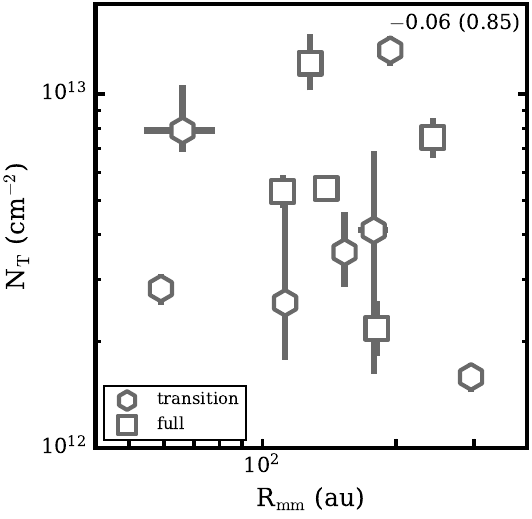}
\caption{CS column densities versus the millimeter dust disk size. Pearson’s correlation coefficient and the associated p-value is shown in the upper right.}
\label{fig:correlation_Rmm_NT}
\end{figure}

Here, we show additional correlation plots with CS T$_{\rm{rot}}$ and N$_{\rm{T}}$ versus the millimeter dust cavity sizes (Figure \ref{fig:correlation_cavity_size}) and the gas-phase C/O ratios (Figure \ref{fig:correlation_CO_Ratio}). We also show the CS N$_{\rm{T}}$ versus the size of the millimeter dust disk size in Figure \ref{fig:correlation_Rmm_NT}.

\clearpage

\bibliography{CS_survey}{}
\bibliographystyle{aasjournalv7}



\end{document}